\pdfoutput=1

\documentclass[12pt]{article}
\usepackage{epsfig}
\usepackage{hhline}
\usepackage{amsmath}
\usepackage{amssymb}
\usepackage{color}
\usepackage{xspace}
\usepackage{paralist} % compactitem
\usepackage{caption}
\usepackage{bm} % 'bold' math symbols (better use \usepackage{newtxtext,newtxmath}, if available on latex distribution)
\usepackage{acronym}

\usepackage{hyperref} % has to be last package loaded
\hypersetup{colorlinks=true, urlcolor=blue}
\usepackage{cite} % DB: enable also clickable references (must be loaded after hyperref)
\hypersetup{
  citecolor=[rgb]{0.,0.,0.5},
  linkcolor=[rgb]{0.6,0.,0.0},
  urlcolor=[rgb]{0.,0.,0.5}
  }

\usepackage{parskip}
\usepackage[utf8]{inputenc}
\newlength{\dinwidth}
\newlength{\dinmargin}
\newcommand{\NNLOJET}{NNLO\protect\scalebox{0.8}{JET}\xspace}

\begin{document}
\pagestyle{empty}

\newcommand{\GeVsq}{\ensuremath{\mathrm{GeV}^2} }
\newcommand{\GeV}{\ensuremath{\mathrm{GeV}} }
\newcommand{\pt}{\ensuremath{P_{T}}}
\newcommand{\PP}{\ensuremath{\mathcal{P}}}
\newcommand{\Qsq}{\ensuremath{Q^{2}}}

%% unfolding
\newcommand{\chisq}{\ensuremath{\chi^{2}}}
\newcommand{\ndf}{\ensuremath{n_{\rm dof}}}
\newcommand{\etajet}{\ensuremath{\eta_{\rm lab}^{\rm jet}}}
\newcommand{\ptjet}{\ensuremath{P_{\rm T}^{\rm jet}}}
\newcommand{\meanpt}{\ensuremath{\langle P_{\rm T} \rangle}}
\newcommand{\etalab}{\ensuremath{\eta_{\rm lab}^{\rm jet}}}
\newcommand{\Mjj}{\ensuremath{m_{12}}}
\newcommand{\meanptdi}{\ensuremath{\langle P_{\mathrm{T}} \rangle_{2}}\xspace}
\newcommand{\meanpttri}{\ensuremath{\langle P_{\mathrm{T}} \rangle_{3}}\xspace}
\newcommand{\mz}{\ensuremath{m_{\rm Z}}\xspace}
\newcommand{\as}{\ensuremath{\alpha_{\rm s}}\xspace}
\newcommand{\asmz}{\ensuremath{\as(\mz)}\xspace}
\newcommand{\asmzPDF}{\ensuremath{\as^{\rm PDF}(\mz)}\xspace}
\newcommand{\asmzf}{\ensuremath{\as^{\Gamma}(\mz)}\xspace}
\newcommand{\tilmu}{\ensuremath{\tilde{\mu}}\xspace}
\newcommand{\etal}{{\it{et al.}}}
\newcommand{\mur}{\ensuremath{\mu_{\rm R}}\xspace}
\newcommand{\muf}{\ensuremath{\mu_{\rm F}}\xspace}
\newcommand{\mup}{\ensuremath{\mu_{0}}\xspace}
\newcommand{\murf}{\ensuremath{\mu_{\rm R/F}}\xspace}
\newcommand{\asmur}{\ensuremath{\alpha_{\rm s}(\mur)}\xspace}
\newcommand{\chad}{\ensuremath{c_{\rm had}}\xspace}
\newcommand{\ord}{\ensuremath{\mathcal{O}}\xspace}
\newcommand{\PDFasResult}{0.1147}
\newcommand{\HonePDF}{H1PDF2017\,{\protect\scalebox{0.8}{[NNLO]}}}

\renewcommand{\contentsname}{Content \footnotesize (only for editorial purposes)}

\hyphenation{QCDNUM}
%%%%%%%%%%%%%%%%%%%%%%%%%%%%%%%%%%%%%%% title page %%%%%%%%%%%%%%%%%%%%%%%%%%%%%%%%%%%%%%%%
\begin{titlepage}
\noindent
\begin{flushleft}
{\tt DESY 17-137    \hfill    ISSN 0418-9833} \\
{\tt September 2017}                  \\
\end{flushleft}

\vspace{1.0cm}
\begin{center}
\begin{Large}

{\bf
  Determination of the strong coupling constant \boldmath{\asmz}
  in next-to-next-to-leading order QCD using H1 jet cross section
  measurements
}

\vspace{1.0cm}

H1 Collaboration

\end{Large}
\end{center}

\vspace{1.0cm}

\begin{abstract}
\noindent
The strong coupling constant \as\ is determined from inclusive jet and
dijet cross sections in neutral-current deep-inelastic $ep$ scattering
(DIS) measured at HERA by the H1 collaboration  using
next-to-next-to-leading order (NNLO) QCD predictions.
The dependence of the NNLO predictions and of the
resulting value of \asmz\ at the $Z$-boson mass $m_Z$ are studied  as
a function of the choice of the
renormalisation and factorisation scales.
Using inclusive jet and dijet data together, the
strong coupling constant is determined to be
$\asmz=0.1166\,(19)_{\rm exp}\,(24)_{\rm th}$.
% PDF+as fit
Complementary, \asmz\ is determined together with parton
distribution functions of the proton (PDFs) from jet and inclusive DIS
data measured by the H1 experiment.
The value $\asmz=\PDFasResult\,(25)_{\rm tot}$ obtained is consistent
with the determination from jet data alone.
The impact of the jet data on the PDFs is studied.
% running
The running of the strong coupling is tested at
different values of the renormalisation scale and the results are
found to be in agreement with expectations.  
\noindent
\end{abstract}

\vspace{1.cm}
\begin{center}
\footnotesize Dedicated to the memory of our dear friends and colleagues Vitaliy
Dodonov and Yakov Vazdik.
\end{center}
\vspace{1.cm}

\begin{center} Published in EPJ C \end{center}

\vspace{1.cm}
%\begin{abstract}
\noindent
{\bf Erratum.}
An implementation error in the NNLO predictions was found~\cite{Currie:2017tpe} which changes the
numerical values of the predictions and the resulting values of the fits.
In the present document, all values, the results and the dicsussion were corrected accordingly.
%\end{abstract}
\begin{center} Erratum submitted to EPJ C \end{center}

\end{titlepage}
%% ------------------------------------------------------------------------------ %%

%% ------------------------------------------------------------------------------ %%
\begin{flushleft}
\noindent
V.~Andreev$^{19}$,             %LPI -PD        8/88            Andreev             
A.~Baghdasaryan$^{31}$,        %YERE-PD        09/03           Baghdasaryana       
K.~Begzsuren$^{28}$,           %ULBA-PD        04/06           Begzsuren           
A.~Belousov$^{19}$,            %LPI -HON       12/12           Belousov            
V.~Bertone$^{46,47,\star}$,
A.~Bolz$^{12}$,                %HDB1-ST        06/15           Bolz                
V.~Boudry$^{22}$,              %ECPL-PD        1/93            Boudry              
G.~Brandt$^{41}$,              %GOET-PD        03/07           Brandt              
V.~Brisson$^{21}$,             %ORSA-HON       01/12           Brisson             
D.~Britzger$^{12}$
% ...

%\newpage

%-- H1 Institutes 
\bigskip{\it
\noindent
% ...
$ ^{12}$ Physikalisches Institut, Universit\"at Heidelberg,
          Heidelberg, Germany$^{ a}$ \\
% ...
$ ^{19}$ Lebedev Physical Institute, Moscow, Russia \\
% ...
$ ^{21}$ LAL, Universit\'e Paris-Sud, CNRS/IN2P3, Orsay, France \\
 $ ^{22}$ LLR, Ecole Polytechnique, CNRS/IN2P3, Palaiseau, France \\
% ...
$ ^{28}$ Institute of Physics and Technology of the Mongolian
          Academy of Sciences, Ulaanbaatar, Mongolia \\
% ...
$ ^{31}$ Yerevan Physics Institute, Yerevan, Armenia \\
% ...
$ ^{41}$ Now at II. Physikalisches Institut, Universit\"at G\"ottingen,
          G\"ottingen, Germany \\
% ...
$ ^{46}$ Vrije University, Department of Physics and Astronomy
         De Boelelaan 1081, Amsterdam, Netherlands \\
 $ ^{47}$ National Institute for Subatomic Physics (NIKHEF)
         Science Park 105, Amsterdam, Netherlands \\
% ...
$ ^{\star}$ Contributing author providing theory predictions or infrastructure \\

\bigskip\noindent
 $ ^a$ Supported by the Bundesministerium f\"ur Bildung und Forschung, FRG,
      under contract numbers 05H09GUF, 05H09VHC, 05H09VHF,  05H16PEA \\
% ...
}
\end{flushleft}
%
% Please not that the author list may need re-formatting.
%% ------------------------------------------------------------------------------ %%

%% ------------------------------------------------------------------------------ %%
\clearpage
%\tableofcontents
%\clearpage
%% ------------------------------------------------------------------------------ %%

\pagestyle{plain} %% page numbers

%%%%%%%%%%%%%%%%%%%%%%%%%%%%%%%%%%%%%%%%%%%%%%%%%%%%%%%%%%%%
%                    Introduction
%%%%%%%%%%%%%%%%%%%%%%%%%%%%%%%%%%%%%%%%%%%%%%%%%%%%%%%%%%%%
\section{Introduction}
% strong coupling constant
The strong coupling constant is one of the least well
known parameters of the Standard Model of particle physics (SM) and a
precise knowledge of this coupling is crucial for
precision measurements, consistency tests of the SM and searches for
physics beyond the SM.  
It has been determined in a large variety of processes and using different
techniques~\cite{Dissertori:2015tfa,Olive:2016xmw}.
% jets used for alpha_s
Jet production in the Breit frame~\cite{Streng:1979pv} in neutral-current deep-inelastic
$ep$ scattering (NC DIS) is directly sensitive to the strong coupling and
has a clean experimental signature with sizable cross sections.
It is thus ideally suited for the precision  determination
of the strong coupling constant \asmz\ at the $Z$-boson mass $m_Z$.

% NNLO calculations first
Cross section 
predictions for inclusive jet and dijet production in NC DIS are
obtained within the framework of perturbative QCD (pQCD)~\cite{ISBN:9780521581899},
where for the past 25~years only next-to-leading order (NLO)
calculations have been available~\cite{Graudenz:1990ej,Graudenz:1990vk}.
Continuous developments enabled the advancement of these 
calculations~\cite{Kosower:1997zr,GehrmannDeRidder:2005cm,Daleo:2009yj,Currie:2013vh},
and next-to-next-to-leading order (NNLO) predictions for jet
production in DIS~\cite{Currie:2016ytq,Currie:2017tpe} and
hadron-hadron collisions~\cite{Currie:2016bfm,Currie:2017eqf} have
become available recently.
The theoretical uncertainties of the NNLO predictions are
substantially reduced compared to those of the NLO predictions.
It is observed~\cite{Currie:2016ytq,Andreev:2016tgi,Currie:2017tpe}
that the NNLO predictions and the
current experimental data are of comparable precision
for large parts of the measured phase space. 

% jet cross section measurements
Measurements of inclusive jet and dijet cross sections in NC DIS have been
performed at HERA by the 
H1~\cite{Adloff:1998vc,Adloff:2000tq,Adloff:2002ew,Aktas:2003ja,Aktas:2004px,Aktas:2007aa,Aaron:2009vs,Aaron:2010ac,Andreev:2014wwa,Andreev:2016tgi}
and
ZEUS~\cite{Breitweg:2000sv,Chekanov:2001fw,Chekanov:2002be,Chekanov:2004hz,Chekanov:2006xr,Chekanov:2006yc,Abramowicz:2010cka,Abramowicz:2010ke}
collaborations during different data taking periods and for 
different centre-of-mass energies.
In general, the predictions in pQCD provide a good description of these
data. 

% alpha_s from DIS jets
The strong coupling constant has been determined from jet cross sections
in DIS at NLO
accuracy~\cite{Ahmed:1994ib,Adloff:1998ss,Adloff:1998kh,Adloff:2000tq,Chekanov:2002be,Chekanov:2006yc,Aktas:2007aa,Aaron:2009vs,Aaron:2010ac,Andreev:2014wwa,Andreev:2016tgi}
and the precision of \asmz\ of these determinations is
typically limited by the scale uncertainty of the NLO calculations.
Only recently an \as\ determination was performed using inclusive jet
cross sections, where NLO calculations have been supplemented with
contributions beyond NLO in the threshold
resummation formalism, and a moderate reduction of the scale
uncertainty was achieved~\cite{Biekotter:2015nra}.

% jets at LHC and Tevatron
% Tevatron, LHC jets
% e+e- 3- and 4-jet rates (jade, LEP)
% e+e- event shapes
% ep event shapes
Measurements of jet production cross sections in processes other than
NC DIS, such as
photoproduction~\cite{Abramowicz:2012jz,Klasen:2013cba} or in
$e^+e^-$~\cite{Abbiendi:2005vd,Schieck:2006tc,Dissertori:2007xa,Dissertori:2009qa,Dissertori:2009ik,Schieck:2012mp},
$p\bar{p}$~\cite{Affolder:2001hn,Abazov:2009nc,Abazov:2012lua} 
and $pp$
collisions~\cite{Malaescu:2012ts,Chatrchyan:2013txa,Khachatryan:2014waa,CMS:2014mna,Khachatryan:2016mlc},
have also been employed for the determination of the strong coupling constant.
The corresponding predictions were at NLO accuracy in most cases,
possibly supplemented with 2-loop threshold corrections or matched
with next-to-leading logarithmic approximations (NLLA). An exception are 3-jet
observables in $e^+e^-$ collisions using predictions in NNLO accuracy~\cite{Dissertori:2009qa}, which are also
matched to NLLA contributions~\cite{Dissertori:2009ik,Schieck:2012mp}.
In contrast to variables such as hadronic event shape
observables~\cite{Aktas:2005tz,OPAL:2011aa}
where only limited regions of the corresponding distributions are
described by fixed order pQCD calculations, jet
observables such as their transverse momenta typically are well
described by such calculations over the full experimentally
accessible range.

% --- PDF+alpha_s fits
The presence of a proton in the initial state in lepton-hadron or
hadron-hadron collisions complicates the determination of \asmz\
and therefore \asmz\ is often determined together with
parton distribution functions of the proton (PDFs).
Such simultaneous determinations of \asmz\ and PDFs were performed using
jet cross sections in DIS~\cite{Adloff:2000tq,Chekanov:2005nn,Abramowicz:2015mha} or jet
cross sections at either the LHC or
Tevatron~\cite{Ball:2011us,Harland-Lang:2014zoa,Sirunyan:2017skj,Khachatryan:2014waa,Khachatryan:2016mlc}. 
However, the absence of full NNLO corrections for jet production cross sections
limited the theoretical precision of these approaches.

% --- what this paper is about
% --- outline of the article
This article presents the first determination of \asmz\ making use of the
recent calculations of jet production at
NNLO~\cite{Currie:2016ytq,Currie:2016bfm,Currie:2017tpe,Currie:2017eqf}.
These calculations are also used in this paper for the first time for the
determination of PDFs.
The jet cross section calculations are performed using the program
\NNLOJET~\cite{Ridder:2016nkl,Currie:2017tpe,Currie:2016ytq}.

Two strategies for the extraction of \asmz\ are investigated.
First, described in section~\ref{sec:asfit}, the value of \asmz\ is
determined in NNLO from inclusive jet and dijet cross
sections~\cite{Adloff:2000tq,Aaron:2010ac,Aktas:2007aa,Andreev:2014wwa,Andreev:2016tgi}
using pre-determined PDFs as input. 
In a second approach described in section~\ref{sec:pdfasfit}, the
value of \asmz is determined together with the PDFs.
This approach is denoted as `PDF+\as-fit' in the following and
uses inclusive DIS
data~\cite{Adloff:1999ah,Adloff:2000qj,Adloff:2003uh,Collaboration:2010ry,Aaron:2012qi,Aaron:2012kn}
in addition to normalised jet
cross section data~\cite{Aktas:2007aa,Andreev:2014wwa,Andreev:2016tgi}, both measured by the H1
experiment~\cite{Abt:1996hi,Abt:1996xv,Appuhn:1996na,Andrieu:1993kh}.

%%%%%%%%%%%%%%%%%%%%%%%%%%%%%%%%%%%%%%%%%%%%%%%%%%%%%%%%%%%%
%                    Data
%%%%%%%%%%%%%%%%%%%%%%%%%%%%%%%%%%%%%%%%%%%%%%%%%%%%%%%%%%%%
%\clearpage
\section{Cross section measurements}
\label{sec:CrossSections}
For the present analysis, measurements of jet cross sections and inclusive DIS cross sections in
lepton-proton collisions performed by the H1 experiment at HERA are exploited.

\paragraph{Jet cross sections}
Cross sections for jet production in lepton-proton collisions have
been measured by H1 at two different centre-of-mass energies using data
from different periods of data taking.  
In the present analysis, inclusive jet and dijet cross 
sections measured in the range of negative four-momentum transfer squared
$5<\Qsq<15\,000\,\GeVsq$ and inelasticities $0.2<y<0.7$ are considered.
An overview of the individual measurements~\cite{Adloff:2000tq,Aaron:2010ac,Aktas:2007aa,Andreev:2014wwa,Andreev:2016tgi} is given in
table~\ref{tab:datasets}.
\begin{table}[tbhp]
  \footnotesize
  %\scriptsize
  \begin{center}
    \begin{tabular}{cccccc}
%      \multicolumn{6}{c}{\bf Kinematic range of H1 jet data} \\
      \hline
      \multicolumn{1}{c}{Data set} & $\sqrt{s}$ & $\mathcal{L}$ & DIS kinematic &  Inclusive jets &  Dijets   \\  
      \multicolumn{1}{c}{[ref.]}  & $[\GeV]$   & $[{\rm pb}^{-1}]$  &  range        &                 &   $n_{\rm jets}\ge2 $  \\   
      \hline
      $300\,\GeV$ & 300 & 33& $150<\Qsq<5000\,\GeVsq$  & $7<\ptjet<50\,\GeV$ & $\ptjet>7\,\GeV$  \\
      \cite{Adloff:2000tq} &          & & $0.2<y<0.6$             &            & $8.5<\meanpt<35\,\GeV$    \\
      \hline
      HERA-I    & 319  &43.5 & $5<\Qsq<100\,\GeVsq$   &   $5<\ptjet<80\,\GeV$ & $\ptjet>5\,\GeV$  \\
\cite{Aaron:2010ac}  &      &     & $0.2<y<0.7$                &                      & $5<\meanpt<80\,\GeV$  \\
                &           &     &                            &                      & $\Mjj>18\,\GeV$  \\
                &           &     &                            &                      & $(\meanpt>7\,\GeV)^*$ \\
      \hline
      HERA-I    & 319  &65.4 & $150<\Qsq<15000\,\GeVsq$   &   $5<\ptjet<50\,\GeV$ & $-$  \\
\cite{Aktas:2007aa} &  &     & $0.2<y<0.7$             &                      &  \\
      \hline
      HERA-II   & 319  & 290& $5.5<\Qsq<80\,\GeVsq$        & $4.5<\ptjet<50\,\GeV$ & $\ptjet>4\,\GeV$  \\
\cite{Andreev:2016tgi}&& & $0.2<y<0.6$                &                      & $5<\meanpt<50\,\GeV$  \\
      \hline
      HERA-II   & 319  & 351& $150<\Qsq<15000\,\GeVsq$     &   $5<\ptjet<50\,\GeV$ & $5<\ptjet<50\,\GeV$  \\
 \cite{Andreev:2014wwa,Andreev:2016tgi}               &           & & $0.2<y<0.7$                &                      & $7<\meanpt<50\,\GeV$  \\
                &           & &                            &                      & $\Mjj>16\,\GeV$  \\
      \hline
    \end{tabular}
    \caption{
      Summary of the kinematic ranges of the studied inclusive
      jet and dijet data sets. 
      The $ep$ centre-of-mass energy $\sqrt{s}$ and the integrated luminosity $\mathcal{L}$
      are shown. Kinematic restrictions are made on the negative four-momentum
      transfer squared \Qsq, the inelasticity $y$ and the jet transverse momenta
      \ptjet\ as indicated. Common to all data sets is a requirement on the
      pseudorapidity of the jets, $-1<\etalab<2.5$, not shown in the
      table. Dijet events are defined by extra cuts or on the average jet
      transverse momentum \meanpt\ or the invariant mass of the two leading
      jets \Mjj. The asterisk denotes a cut not present in the original
      work~\cite{Aaron:2010ac} but imposed for the present analysis.
    }
    \label{tab:datasets}
    \end{center}
\end{table}
Common to all data, jets are defined in the Breit frame~\cite{Streng:1979pv} using the
$k_t$ clustering algorithm~\cite{Ellis:1993tq} with a resolution
parameter $R=1$. 
The jet four-vectors are restricted to the pseudorapidity range
$-1<\etalab<2.5$ in the laboratory frame. 
The data sets `$300\,\GeV$', `HERA-I' and `HERA-II' correspond to different
data taking periods and are subdivided into two kinematic ranges, the low-\Qsq\ ($\Qsq\lesssim100\,\GeVsq$) and high-\Qsq\ 
($\Qsq\gtrsim150\,\GeVsq$) domains, 
where different components of the H1 detector were used
for the measurement of the scattered lepton.

The inclusive jet cross sections are measured double-differentially as
functions of \Qsq\ and the jet transverse momentum in the Breit frame,
\ptjet, where the phase space is constrained by
\Qsq, $y$, \etalab\ and \ptjet, as specified in
table~\ref{tab:datasets}.

For dijets at least two jets must be identified 
in the \etalab\ range  above the relevant \ptjet\ threshold.
The double-differential dijet cross sections are measured as functions
of \Qsq\ and the average transverse momentum 
of the two leading jets, $\meanpt=(P_{\rm T}^{\rm jet1} + P_{\rm T}^{\rm jet2})/2$.
In order to avoid regions of phase space where the
predictions exhibit an enhanced infrared sensitivity~\cite{Duprel:1999wz,Potter:1999gg}, the phase
space definitions impose asymmetric cuts on the transverse momenta of
the two leading jets~\cite{Currie:2017tpe}. 
Such an asymmetric cut may also be obtained by choosing
\meanpt\ larger than the minimum \ptjet.
For this reason, data points with $\meanpt < 7\,\GeV$ are excluded
from the HERA-I low-\Qsq\ data set (table~\ref{tab:datasets}).

Data from different periods and \Qsq\ ranges are statistically independent, 
whereas dijet and inclusive jet data of the same data set are
statistically correlated. 
These correlations have been determined for the HERA-II data
sets~\cite{Andreev:2014wwa,Andreev:2016tgi}.
Different data sets, as well as inclusive jet and dijet data of the
same data set, may furthermore share individual sources of experimental
uncertainties~\cite{Andreev:2016tgi,Abramowicz:2015mha} and thus
correlations are present for all data points considered.

\paragraph{Normalised jet cross sections}
The more recent data sets~\cite{Aktas:2007aa,Andreev:2014wwa,Andreev:2016tgi}
also include measurements where the jet cross sections are
normalised to the inclusive NC DIS
cross section of the respective \Qsq\ interval, as indicated in
table~\ref{tab:normdata}.
%A summary of these
%measurements is listed in table~\ref{tab:normdata}.
\begin{table}[tbhp]
  \footnotesize 
  %\scriptsize
  \begin{center}
    \begin{tabular}{lcccccc}
%      \multicolumn{7}{c}{\bf H1 absolute and normalised jet cross sections} \\
      \hline
      \multicolumn{1}{c}{Data set} & \Qsq\ domain & Inclusive & Dijets
      & Normalised    & Normalised  & Stat. corr. \\  
      \multicolumn{1}{c}{[ref.]}   &              & jets &       &
      inclusive jets &   dijets  & between samples    \\
      \hline
      $300\,\GeV$~\cite{Adloff:2000tq} & high-\Qsq & \checkmark  & \checkmark & --  & --  & --  \\
      HERA-I~\cite{Aaron:2010ac}       & low-\Qsq  & \checkmark  & \checkmark & --  & --  & -- \\
      HERA-I~\cite{Aktas:2007aa}       & high-\Qsq & \checkmark  & --         & \checkmark & --  & -- \\
      HERA-II~\cite{Andreev:2016tgi}   & low-\Qsq  & \checkmark  & \checkmark & \checkmark & \checkmark & \checkmark  \\
      HERA-II~ \cite{Andreev:2014wwa,Andreev:2016tgi}
                                       & high-\Qsq & \checkmark & \checkmark & \checkmark & \checkmark & \checkmark  \\
      \hline
    \end{tabular}
    \caption{
      H1 jet cross section
      measurements. Normalised dijet cross sections and statistical
      correlations between inclusive and dijet measurements are available
      only for the most recent measurements~\cite{Andreev:2014wwa,Andreev:2016tgi}.
    }
    \label{tab:normdata}
    \end{center}
\end{table}
Correlations of systematic and statistical uncertainties
partially cancel for the ratio of jet cross sections and inclusive NC
DIS cross sections. Therefore, normalised jet cross sections are ideally suited for
studies together with inclusive NC DIS data.

% --- inclusive DIS data
\paragraph{Inclusive DIS cross sections}
In order to constrain the parameters of the PDFs in the PDF+\as-fit,
polarised and unpolarised inclusive NC and CC (charged current) DIS cross
sections~\cite{Adloff:1999ah,Adloff:2000qj,Adloff:2003uh,Collaboration:2010ry,Aaron:2012qi,Aaron:2012kn}
measured by the H1 experiment are used in addition.
Data taken during different data taking periods and with different
centre-of-mass energies are considered and a summary of these
measurements is given in table~\ref{tab:DISdata}.
\begin{table}[tbhp]
  \footnotesize
  \begin{center}
    \begin{tabular}{lcccccc}
      \hline
      \multicolumn{1}{c}{Data set} & Lepton & $\sqrt{s}$ & \Qsq\ range &  NC cross    & CC cross    & Lepton beam \\  
      \multicolumn{1}{c}{[ref.]}   & type & $[\GeV]$    & $[\GeVsq]$ &  sections & sections & polarisation \\
      \hline
      Combined low-\Qsq~\cite{Collaboration:2010ry}  & $e^+$ & 301,319 & (0.5) 12 -- 150 & \checkmark & -- &   --  \\
      Combined low-$E_p$~\cite{Collaboration:2010ry} & $e^+$ & 225,252 & (1.5) 12 --  90 & \checkmark & -- &   --  \\
      94 -- 97~\cite{Adloff:1999ah}              & $e^+$ & 301 &  150 -- 30\,000 & \checkmark  & \checkmark  &   -- \\
      98 -- 99~\cite{Adloff:2000qj,Adloff:2003uh}& $e^-$ & 319 &  150 -- 30\,000 & \checkmark  & \checkmark  &   -- \\
      99 -- 00~\cite{Adloff:2003uh}              & $e^+$ & 319 &  150 -- 30\,000 & \checkmark  & \checkmark  &   -- \\
      HERA-II~\cite{Aaron:2012qi}                & $e^+$ & 319 &  120 -- 30\,000 & \checkmark  & \checkmark  &   \checkmark \\
      HERA-II~\cite{Aaron:2012qi}                & $e^-$ & 319 &  120 -- 50\,000 & \checkmark  & \checkmark  &   \checkmark \\
      \hline
    \end{tabular}
    \caption{
      Summary of the inclusive NC and CC DIS data sets. 
      The lepton type, the $ep$ centre-of-mass energy $\sqrt{s}$ 
      and the considered \Qsq\ range are shown. 
      The numbers in parenthesis show the whole kinematic range of
      the data prior to applying the $Q^2$ cut specific for this analysis.
      The check-marks indicate the available measurements.
      The last column indicates cross sections determined with longitudinally
      polarised leptons.
    }
    \label{tab:DISdata}
    \end{center}
\end{table}
This data sample is identical to the one used in the H1PDF2012 PDF
fit~\cite{Aaron:2012qi}, where correlations of experimental uncertainties
have been quantified.
Inclusive DIS and jet cross sections are statistically and experimentally
correlated. 
These correlations are taken into account by using normalised jet cross sections.

%%%%%%%%%%%%%%%%%%%%%%%%%%%%%%%%%%%%%%%%%%%%%%%%%%%%%%%%%%%%
%                    as-fit
%%%%%%%%%%%%%%%%%%%%%%%%%%%%%%%%%%%%%%%%%%%%%%%%%%%%%%%%%%%%
\begin{boldmath}
\section{Determination of \asmz\ from H1 jet cross sections}
\label{sec:asfit}
\end{boldmath}
The strong coupling constant \asmz\ is determined from inclusive jet and
dijet cross sections in NC DIS measured by the H1 collaboration and using
NNLO QCD predictions.

%%%%%%%%%%%%%%%%%%%%%%%%%%%%%%%%%%%%%%%%%%%%%%%%%%%%%%%%%%%%
%                    Theory
%%%%%%%%%%%%%%%%%%%%%%%%%%%%%%%%%%%%%%%%%%%%%%%%%%%%%%%%%%%%

\subsection{Predictions}
\label{sec:predictions}
The cross sections for inclusive jet and dijet production for a given
phase space interval~$i$ (for instance a `bin' in the relevant
physical observables) are calculated~\cite{ISBN:9780521581899,ISBN:9789971505653.ch1}
as a convolution in the variable $x$ of the PDFs $f_k$ and perturbatively calculated
partonic cross sections ${\hat\sigma}_{i,k}$, 
\begin{equation}
  \sigma_i = \sum_{k=g,q,\overline{q}} \int dx f_{k} (x,\muf)
  \hat{\sigma}_{i,k}(x,\mur,\muf) \cdot c_{{\rm had},i}~,
  \label{eq:sigma}
\end{equation}
where the sum runs over all parton flavours $k$. 
The calculations depend on the renormalisation scale \mur\ and the factorisation
scale \muf.
The factors $c_{{\rm had},i}$ account for non-perturbative effects 
(hadronisation corrections). 

% --- theory: sensitivity
Both the $f_k$ and the ${\hat\sigma}_{i,k}$ are sensitive to the
strong coupling. The partonic cross sections are given in
terms of the perturbative expansion in orders of
$\asmur$
\begin{equation}
  \hat\sigma_{i,k} = \sum_{n} \as^{n}(\mur)\hat{\sigma}_{i,k}^{(n)}(x,\mur,\muf)\,.
  \label{eq:sigmahat}
\end{equation}
For high \pt\ jet production in the Breit frame the lowest order is
$n=1$.
The hard coefficients $\hat{\sigma}_{i,k}^{(n)}$ are calculated for
the expansion up to $\ord(\as^3)$ 
taking into account properties of the jet algorithm in the integration
over the phase space.
The renormalisation scale dependence (`running') of the coupling
satisfies the renormalisation group equation 
\begin{equation}
  \mur^2\frac{d\as}{d\mur^2} = \beta(\as)\,.
  \label{eq:running}
\end{equation}
The QCD beta-function $\beta$ is known at 4-loop
accuracy~\cite{vanRitbergen:1997va,Czakon:2004bu}.
The strength of the coupling thus may be determined
at an arbitrary scale, which is conventionally chosen to be the  mass of the
$Z$-boson, $\mz=91.1876\,\GeV$~\cite{Olive:2016xmw}.
Here, the calculations are performed in the modified minimal subtraction
($\overline{\rm MS}$) scheme in 3-loop accuracy and using 5 flavors,
$\asmur=\alpha^{(5)}_{\overline{\rm MS}}(\mur)$.

% --- theory: sensitivity PDFs
The PDFs $f_k$ exhibit a dependence on \asmz, which originates
from the factorisation theorem~\cite{ISBN:9789971505653.ch1}. 
This dependence can be schematically expressed as~\cite{Gribov:1972ri,Altarelli:1977zs,Dokshitzer:1977sg} 
\begin{equation}
  \muf^2\frac{d f}{d\muf^2} = \PP(\as)\otimes f
  \label{eq:PDFmuf}
\end{equation}
with $\PP$ being the QCD splitting kernels and the symbol `$\otimes$'
denoting a convolution.
After fixing the $x$-dependence of the PDFs $f_k$ at a scale $\mup$
and setting $\mur=\muf$,
the PDF at any factorisation scale \muf\ is calculated as
\begin{equation}
  f\left(x,\muf,\asmz\right) = \Gamma\left(\muf,\mup,\asmz\right)\otimes f_{\mup}\left(x\right)
  \label{eq:PDF}
\end{equation}
with $\Gamma$ being the evolution kernel which obeys
equation~\ref{eq:PDFmuf}. It is here calculated in NNLO, i.e.\ in
3-loop accuracy~\cite{Moch:2004pa,Vogt:2004mw}, with five active
flavours.

The evolution starting scale is chosen to be $\mup=20\,\GeV$. 
This is a typical scale of the jet data studied.
As a consequence, the influence of the evolution of equation~\ref{eq:PDF}
on the \as\ determination is moderate, because $\muf\approx\mup$. 
The PDFs at that scale are well known,
in particular the quark densities. Moreover, the latter are to a
large extent insensitive to the assumption made on the strong coupling
\asmzPDF\ during their determination, because in leading order QCD
inclusive DIS is independent of \as\.
The gluon density is constrained due to QCD sum-rules and the
precisely known quark densities.
In the vicinity of a scale of $20\,\GeV$ threshold effects from
heavy quarks are not relevant.
The PDFs at $\mup=20\,\GeV$ are provided by the NNPDF3.1 PDF
set~\cite{Ball:2017nwa} which was obtained with a nominal value of
$\asmzPDF=0.118$. The influence of those choices is quantified in
section~\ref{sec:studies}.

%% scale
The scales \mur\ and \muf\ are chosen to be
\begin{equation}
  \mur^2=\muf^2=\Qsq+\pt^2~,
  \label{eq:scale}
\end{equation}
where \pt\ denotes \ptjet\ in the case of inclusive jet cross sections and
\meanpt\ for dijets.
Previously, a variety of different scale definitions have
been employed by
H1~\cite{Ahmed:1994ib,Adloff:1998ss,Adloff:1998st,Adloff:1998kh,Adloff:2000tq,Aaron:2010ac,Aktas:2007aa,Andreev:2014wwa,Andreev:2016tgi},
ZEUS~\cite{Chekanov:2001fw,Chekanov:2002be,Chekanov:2005nn,Chekanov:2006xr,Chekanov:2006yc,Abramowicz:2010cka,Abramowicz:2010ke}
 and elsewhere~\cite{Adloff:1998gg,Duprel:1999wz,Adloff:2002ew,Chekanov:2007aa,Aktas:2007bv,Aaron:2011mp,Andreev:2014yra,Andreev:2015cwa}.
The choice adopted here was already suggested and discussed 
earlier~\cite{Potter:1998fw,Potter:1998jt,Breitweg:2000sv,Chekanov:2004hz}.
Advantages of the scale defined in equation~\ref{eq:scale} are in its simple
functional form and in the fact that it remains non-zero in either
of the kinematical limits $\Qsq\rightarrow0\,\GeVsq$ and $\pt^2\ll\Qsq$.
This is particularly important here, since low- and
high-\Qsq\ domains and a large range in \pt\ are considered.

\begin{figure}[ht]
\begin{center}
   \includegraphics[width=0.99\textwidth]{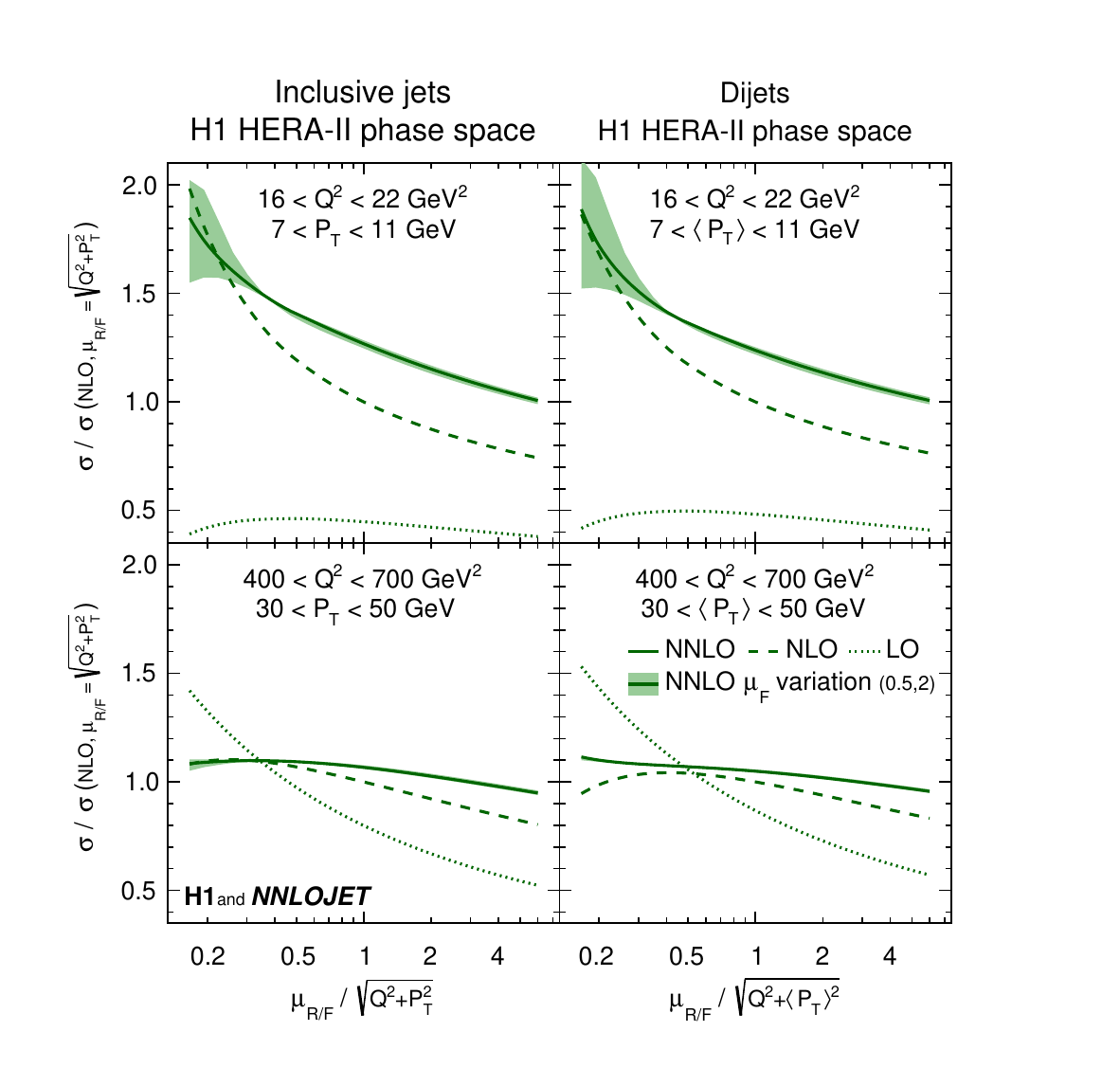}
\end{center}
\caption{
  Relative change of jet cross section as a function of a multiplicative factor applied
  to the renormalisation and factorisation scale for four exemplary
  data points of the HERA-II phase space. 
  The bin definitions are displayed in the respective panels. 
  The left panels show inclusive jet cross sections, and the right panels
  dijet cross sections.
  The full line shows the cross section dependence for the NNLO, the
  dashed line for NLO and the dotted line for LO calculations. For
  better comparison, all calculations are performed with the same PDF
  set (NNPDF3.1 NNLO). 
  For all panels, the cross sections are normalised to the respective NLO cross section with unity scale factor.
  The filled area around the NNLO calculation indicates variations of the factorisation scale by factors of 0.5 and 2 around the chosen value for \mur.
}
\label{fig:sigmascale}
\end{figure}

The inclusive jet and dijet NNLO predictions as a
function of \mur\ and \muf\ %the renormalisation and factorisation
are studied for selected phase space regions in
figure~\ref{fig:sigmascale}.
The dependence on the scale factor is strongest for
cross sections at lower values \mur, i.e.\ lower values of \Qsq\ and \pt. 
The NNLO predictions depend less on the scale factor
than the  NLO predictions. 
Other choices of \mur\ and \muf\ are studied with the \as\ fit in
section~\ref{sec:scalestudy}.

\begin{figure}[ht]
\begin{center}
   \includegraphics[width=0.99\textwidth]{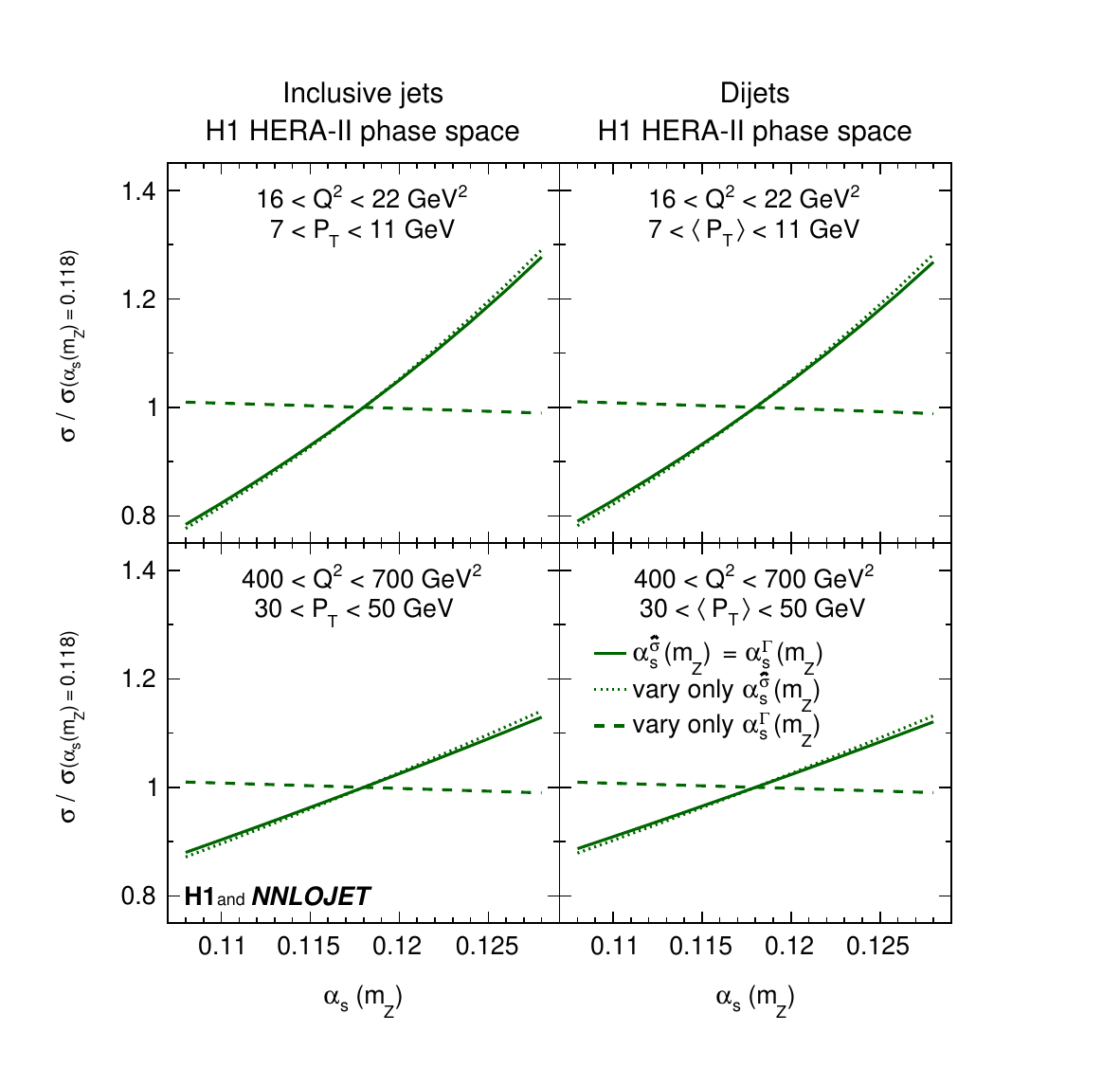}
\end{center}
\caption{
  Relative change of jet cross section as a function of \asmz\ for four exemplary data
  points of the {HERA-II} phase space. The bin definitions are
  displayed in the respective panels. The left panels show inclusive
  jet cross sections, and the right pads dijet cross sections.  
  The full line indicates the cross section dependence as a
  function of \asmz, while the dotted line illustrates the dependence
  where \asmz\ is varied only in the partonic cross sections and the dashed
  line illustrates a variation only in the PDF evolution starting from
  $\mup=20\,\GeV$. The cross sections
  are normalised to the nominal cross section defined with
  $\asmz=0.118$. 
}
\label{fig:sigmaalphas}
\end{figure}

% --- cross section vs. alpha_s plot
The dependence of the inclusive jet and dijet NNLO predictions on \asmz\ is displayed in
figure~\ref{fig:sigmaalphas}, where the two contributions to the
\asmz\ dependence, $\hat\sigma_{ik}$ and $f_k$, are separated.
The predominant sensitivity to \asmz\ arises from $\hat\sigma_{i,k}$.

% ---- Computing packages
The hard coefficients $\hat\sigma^{(n)}_{i,k}$ are calculated using the program
\NNLOJET~\cite{Ridder:2016nkl,Currie:2017tpe,Currie:2016ytq}, which is interfaced to
fastNLO~\cite{Britzger:2012bs} to allow for computationally efficient,
repeated calculations with different values of \asmz, different scale
choices and different PDF sets. 
The PDFs are included in the LHAPDF
package~\cite{Buckley:2014ana}. 
The evolution kernels are calculated using the program APFEL++~\cite{Bertone:2017gds} and
all results are validated with the programs
APFEL~\cite{Bertone:2013vaa} and
QCDNUM~\cite{Botje:2010ay,Botje:2016wbq}. 
The \as\ evolution is calculated using the APFEL++ code and validated with the CRunDec
code~\cite{Schmidt:2012az}, and the
running of the electromagnetic coupling with \Qsq\ is
calculated using the package
EPRC~\cite{Jegerlehner:2008rs,Spiesberger:1995pr}.
The fits are performed using the Alpos fitting framework~\cite{alpos:www}.

%%%%%%%%%%%%%%%%%%%%%%%%%%%%%%%%%%%%%%%%%%%%%%%%%%%%%%%%%%%%
%                    Methodology: as-fit
%%%%%%%%%%%%%%%%%%%%%%%%%%%%%%%%%%%%%%%%%%%%%%%%%%%%%%%%%%%%
%\clearpage
\subsection{Methodology} %\subsection{Methodology for the \boldmath\as\ determination}
%\label{sec:asfit}
The value of the strong coupling constant is determined in a fit of
theory predictions to H1 jet cross sections with a
single free fit parameter.
The goodness-of-fit quantity, which is subject to the minimisation
algorithm, is defined as
\begin{equation}
  \chisq = \sum_{i} \sum_{j}  \left(\log\varsigma_i - \log\sigma_i\right)(V_{\rm exp}
  +V_{\rm had}+V_{\rm PDF})^{-1}_{ij}\left(\log\varsigma_j - \log\sigma_j\right)\,,
  \label{eq:chisq}
\end{equation}
where $\varsigma_i$ are the measurements and $\sigma_i$ the
predictions (equation~\ref{eq:sigma}). The
covariance matrices express the relative uncertainties of the data
($V_{\rm exp}$), hadronisation correction factors ($V_{\rm had}$) and
the PDFs ($V_{\rm PDF}$).
The underlying statistical model is that the logarithm of each
measurement is normal-distributed within its relative uncertainty,
or equivalently the measurements follow log-normal distributions.
The fit value is found using the TMinuit
algorithm~\cite{James:1975dr,Antcheva:2009zz}. 
Correlations of the uncertainties among the different data sets and
running periods are
considered~\cite{Abramowicz:2015mha,Andreev:2016tgi}.
The hadronisation corrections and their uncertainties have been provided
together with the jet cross section measurements~\cite{Adloff:2000tq,Aaron:2010ac,Aktas:2007aa,Andreev:2014wwa,Andreev:2016tgi}.
The PDF uncertainties were provided by the authors of the respective PDF set.

% --- mu > m_b cut
To each data point a representative scale value $\tilmu$ is assigned, which
is calculated from the geometric mean of the bin boundaries
(denoted as `dn' and `up') in \Qsq\ and \pt,
\begin{equation}
  Q^2_{{\rm avg},i} = \sqrt{ Q^2_{{\rm dn},i}  Q^2_{{\rm up},i}}
  ~~{\rm and}~~
  P_{{\rm T,avg},i} = \sqrt{ P_{{\rm T,dn},i} P_{{\rm T,up},i} }\,,
\label{eq:geommean}
\end{equation}
together with the definition of the scales in equation~\eqref{eq:scale} as
\begin{equation}
  \tilde{\mu}^2_i = Q^2_{{\rm avg},i} + P^2_{{\rm T,avg},i}\,.
\label{eq:scaledefinition}
\end{equation}
Effects from heavy quark masses become important at lower
scales, while the NNLO calculations are performed with five massless
quark flavours.
Unless otherwise stated the data are selected with the
condition $\tilmu>2m_b$, with
$m_b=4.5\,\GeV$~\cite{Abramowicz:2015mha} being the mass of the $b$-quark.

% --- uncertainties on fit results
The uncertainty calculated by TMinuit contains the experimental (exp),
hadronisation (had) and PDF uncertainties (PDF). 
The breakdown of the uncertainties into these three components
is obtained from repeated 
fits with $V_{\rm had}$ and/or $V_{\rm PDF}$ set to zero.
Further uncertainties are defined in section~\ref{sec:studies} and will
be denoted as PDFset, PDF\as, and scale uncertainties.
The theory uncertainty (`th') is defined as the quadratic sum of the PDF,
PDFset, PDF\as, hadronisation and scale uncertainties, 
and the `total' uncertainty considers additionally the experimental
uncertainty.

% ----
The value of \asmz\ is determined separately for each individual data
set, for all inclusive jet measurements, for all dijet measurements,
and for all H1 jet data taken together.
The latter is denoted as `H1 jets' in the following.
In the case of fits to `H1 jets', dijet data from the HERA-I running
period however are excluded, since their statistical correlations to the
respective inclusive jet data are not known (table~\ref{tab:normdata}).

%%%%%%%%%%%%%%%%%%%%%%%%%%%%%%%%%%%%%%%%%%%%%%%%%%%%%%%%%%%%
\subsection{Sensitivity of the fit to input parameters}
\label{sec:studies}
\begin{figure}[ht]
\begin{center}
   \includegraphics[width=0.70\textwidth]{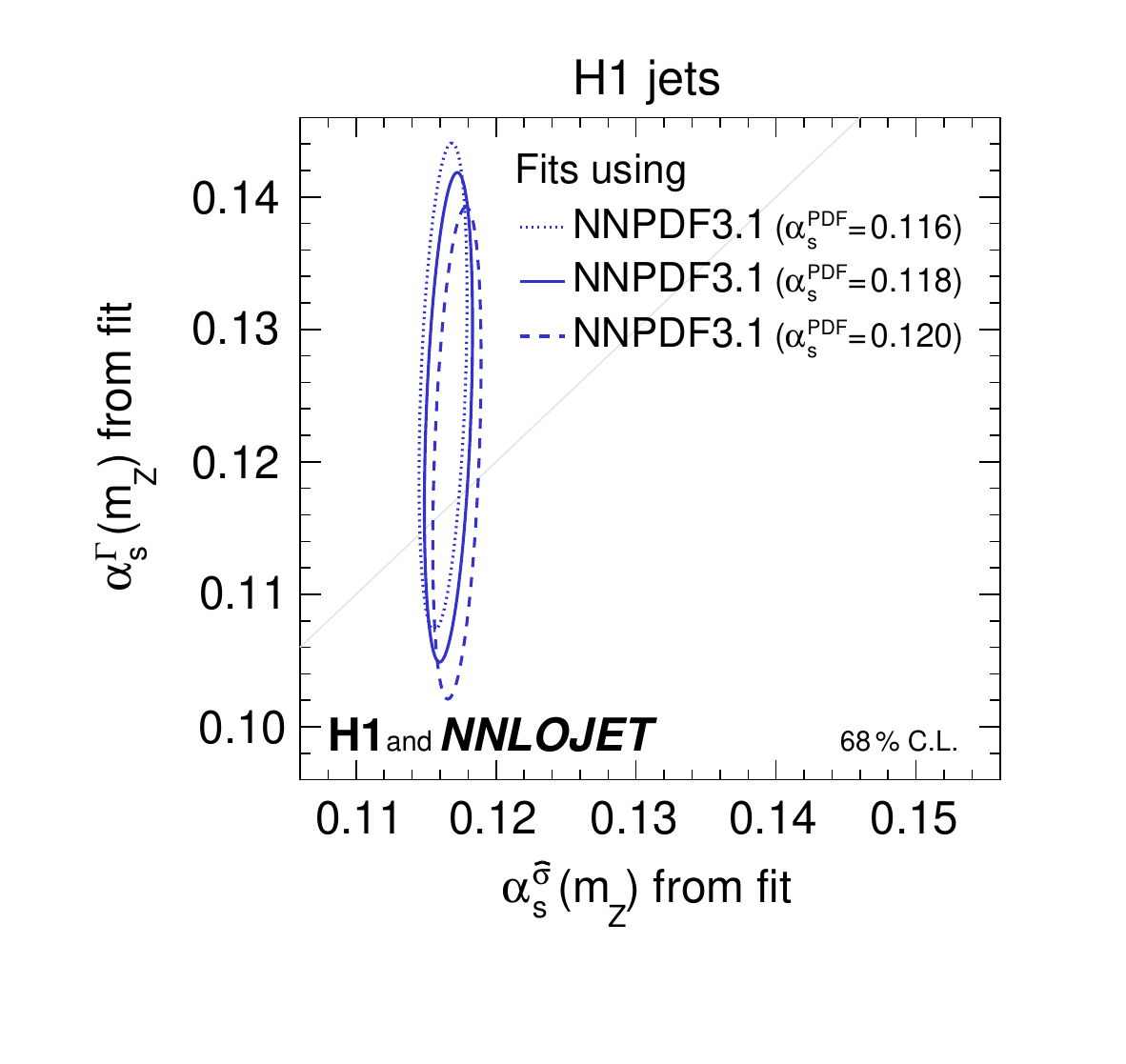}
\end{center}
\caption{
  Results from fits to H1 jets with
  two free fit parameters for \asmz, where the appearances of
  \asmz\ in the PDF evolution $\asmzf$ and in the partonic cross sections
  $\as^\sigma(\mz)$ are identified separately.
  The ellipses display a confidence level of $68\,\%$ including the 
  experimental, hadronisation and PDF uncertainties, and thus the
  lines are calculated for $\Delta\chisq=2.3$.
  The dotted, full and dashed lines indicate the
  contour for $\Delta\chisq=2.3$ using three versions of the NNPDF3.1 set which were
  obtained using values for \asmzPDF\ of 0.116, 0.118 and 0.120,
  respectively.
  The grey straight line corresponds to $\asmzf=\as^\sigma(\mz)$.
}
\label{fig:fit_contours}
\end{figure}

%%%%%%%%%%%%%%%%%%%%%%%%%%%%%%%%%%%%%%%%%%%%%%%%%%%%%%%%%%%%
% --- contours
\paragraph{Sensitivity to \boldmath\asmz}
The sensitivity of the data to \asmz\ and the
consistency of the calculations are investigated by 
performing fits with two free parameters representing the two distinct
appearances of \asmz\ in equation~\eqref{eq:sigma}, i.e.\ in the PDF
evolution, $\asmzf$, and in the partonic cross sections,  
$\as^{\hat\sigma}(m_Z)$. 
The cross sections with the \as\ contributions identified
separately are schematically expressed by
\begin{equation}
  \sigma_i = f \left(\asmzf\right) \otimes  \hat{\sigma}_{i}\left(\as^{\hat\sigma}(m_Z)\right) \cdot c_{{\rm had},i}~,
  \label{eq:sigma2as}
\end{equation}
where $\asmz$ as of equation~\ref{eq:PDF} is denoted as
\asmzf, and \asmz\ as of equation~\ref{eq:sigmahat} is denoted as $\as^{\hat\sigma}(m_Z)$.
The result of such a fit performed for H1 jets is displayed in
figure~\ref{fig:fit_contours}.
Consistency is found for the two fitted values of \asmz, where the
resulting $\asmzf$ tends to be larger than $\as^{\hat\sigma}(m_Z)$.
It is observed that the predominant sensitivity to
\asmz\ arises from the $\hat\sigma_{i,k}$, as was already suggested by the 
jet cross section study (figure~\ref{fig:sigmaalphas}).  
The ellipses obtained using PDFs determined with values \asmzPDF\ of
0.116, 0.118 and 0.120 are consistent with each other. 
In the following, all fits are performed using a single fit parameter \asmz.

%%%%%%%%%%%%%%%%%%%%%%%%%%%%%%%%%%%%%%%%%%%%%%%%%%%%%%%%%%%%
% studies
\paragraph{Dependence on the choice of PDF}
% --- PDFs
Values of \asmz\ are determined for other PDF sets
and for alternative values \asmzPDF.

\begin{figure}[ht]
\begin{center}
   \includegraphics[width=0.99\textwidth]{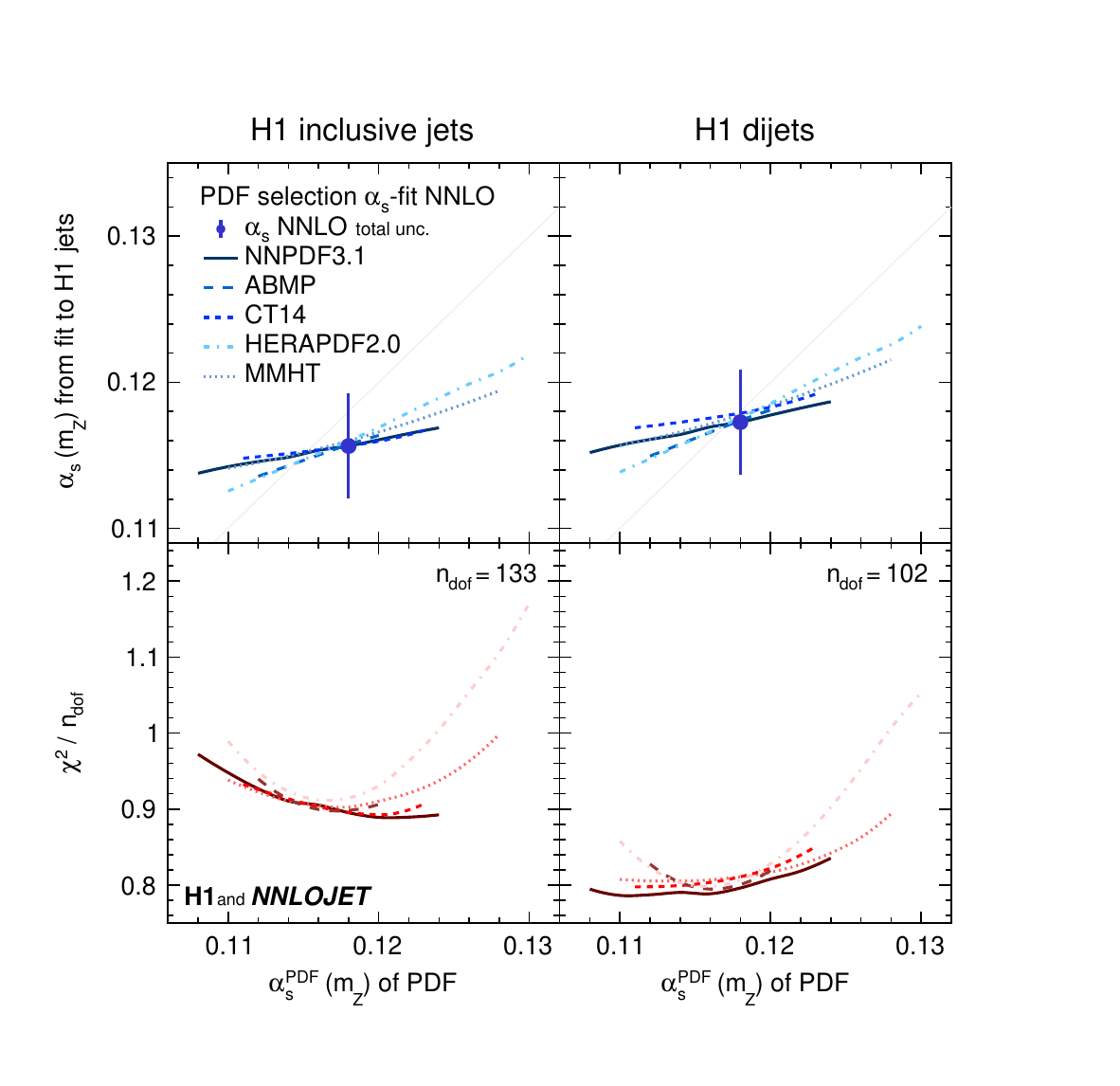}
\end{center}
\caption{
  Dependencies of the fitted values of \asmz\ on the input PDFs for
  separate fits of inclusive jet and dijet data.
  Shown are fits using the ABMP, CT14, HERAPDF2.0, MMHT and NNPDF3.1 PDF sets.
  For each case, the PDFs are available for different input values
  \asmzPDF\ used for the PDF determination, and these values are displayed
  on the  horizontal axis.
  The PDFs are available only for discrete values of \asmzPDF\ and the
  results are connected by smooth lines.
  The lower panel displays the resulting values of $\chisq/\ndf$ of
  the fits.
}
\label{fig:fit_PDF}
\end{figure}

\begin{figure}[ht]
\begin{center}
   \includegraphics[width=0.48\textwidth]{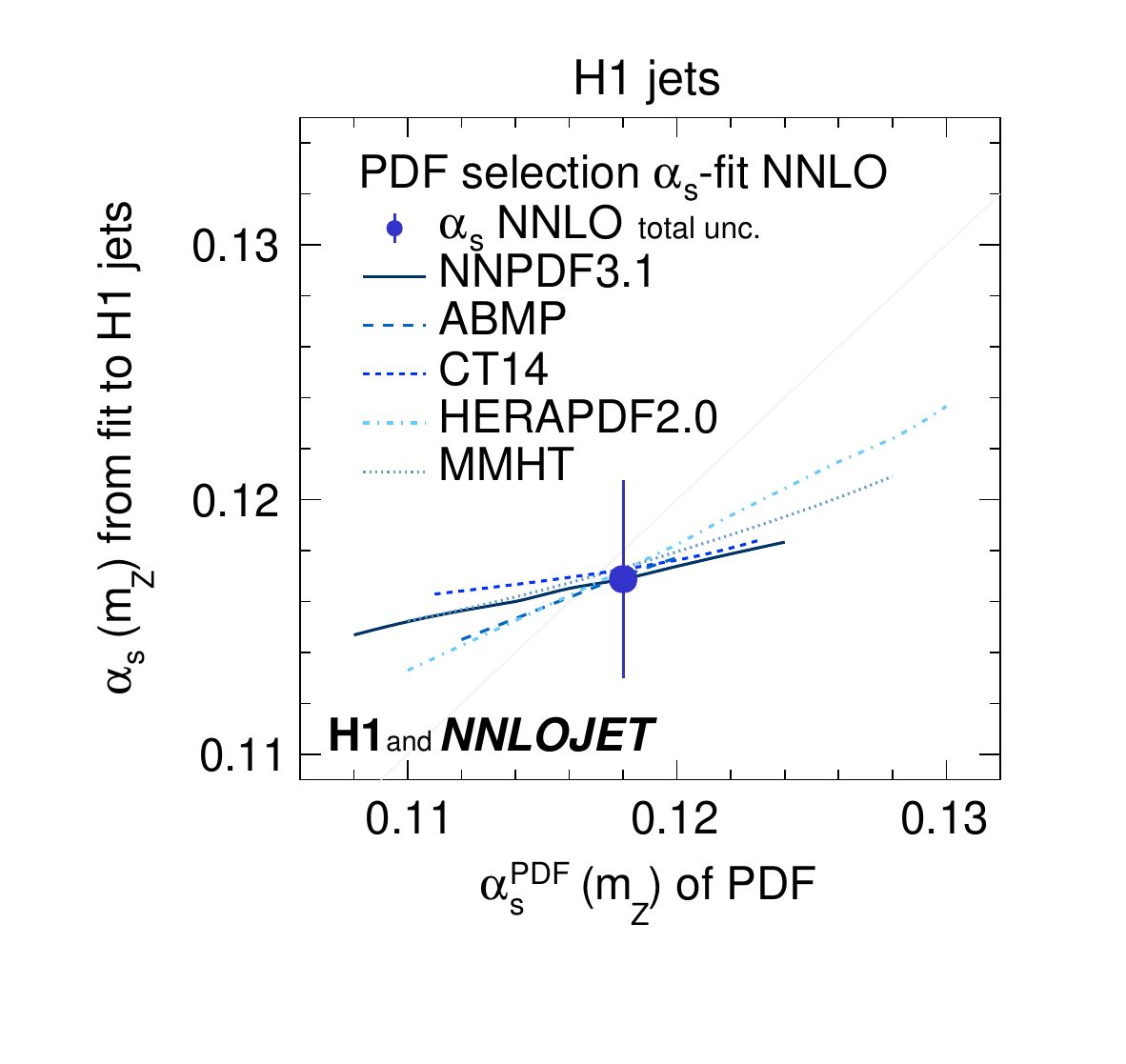}\hfill
   \includegraphics[width=0.48\textwidth]{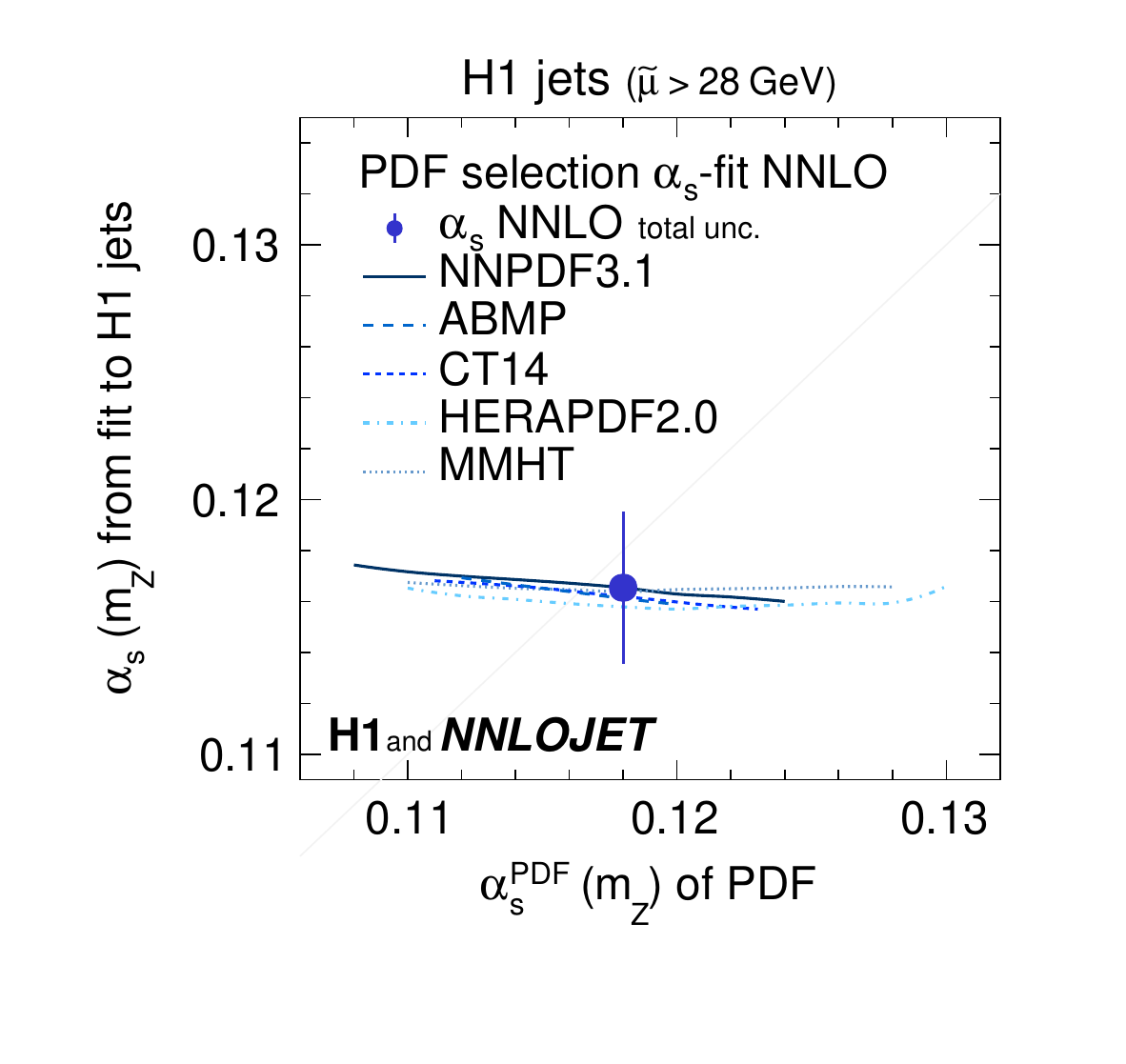}
\end{center}
\caption{
  Dependencies of the fitted values of \asmz\ on the input PDFs for
  the H1 jets fit (left) and the H1 jets fit with $\tilmu>28\,\GeV$
  (right).
  Further details are given in the caption of figure~\ref{fig:fit_PDF}.
}
\label{fig:fit_PDF_MJ}
\end{figure}

The results obtained using different PDFs are displayed in
figure~\ref{fig:fit_PDF} for fits to inclusive jet and 
dijet cross sections, and in figure~\ref{fig:fit_PDF_MJ} for H1 jets. 
In figure~\ref{fig:fit_PDF_MJ} (right) only H1 jets with
$\tilmu>28\,\GeV$ are used.
The predictions using NNPDF3.1, determined with $\asmzPDF=0.118$,
provide good description of the data with \chisq/\ndf\ smaller than unity
(figure~\ref{fig:fit_PDF}),
where \ndf\ denotes the number of data points minus one.
The fitted \asmz\ values are only weakly correlated to the
\asmzPDF\ values employed for the PDF extraction
(figure~\ref{fig:fit_PDF} and~\ref{fig:fit_PDF_MJ}). 
Different PDF sets yield consistent results.
The correlation of \asmzPDF\ and the fitted \asmz\ vanishes when using
only data with $\tilmu>28\,GeV$.

Three PDF related uncertainties are assigned to the fitted \asmz\ results.
The `PDF' uncertainty originates from the data used for the PDF extraction~\cite{Ball:2017nwa}.
A `PDFset' uncertainty is defined as half of the maximum difference
of the results from fits using the
ABMP~\cite{Alekhin:2017kpj}, CT14~\cite{Dulat:2015mca},
HERAPDF2.0~\cite{Abramowicz:2015mha}, MMHT~\cite{Harland-Lang:2014zoa}
or NNPDF3.1 PDF set~\cite{Ball:2017nwa}. % Ball:2014uwa}.
The `PDF\as' uncertainty is defined as the difference of
results from repeated fits using PDFs of the NNPDF3.1 series
determined with \asmzPDF\ values differing by 0.002~\cite{Lai:2010nw}.
This uncertainty can be considered to be uncorrelated to the PDF
uncertainty~\cite{Lai:2010nw,Botje:2011sn}.
The size of the variation includes the NNPDF3.1 PDF set determined with
$\asmzPDF=0.116$,
where \asmzPDF\ is close to the fitted \asmz, in particular
when restricting H1 jets to $\tilmu>28\,\GeV$ (figure~\ref{fig:fit_PDF_MJ}).
The variation of $\mup$ in the range 10 to 90 GeV is also studied
but has negligible effect on the results.

%%%%%%%%%%%%%%%%%%%%%%%%%%%%%%%%%%%%%%%%%%%%%%%%%%%%%%%%%%%%
% studies mur, muf
\paragraph{Scale variants and comparison of NLO and NNLO predictions}
\label{sec:scalestudy}
% --- scale variations
Studies of different choices for \mur\ and \muf\ are commonly used to estimate
contributions of higher orders beyond NNLO.

\begin{figure}[ht]
\begin{center}
   \includegraphics[width=0.99\textwidth]{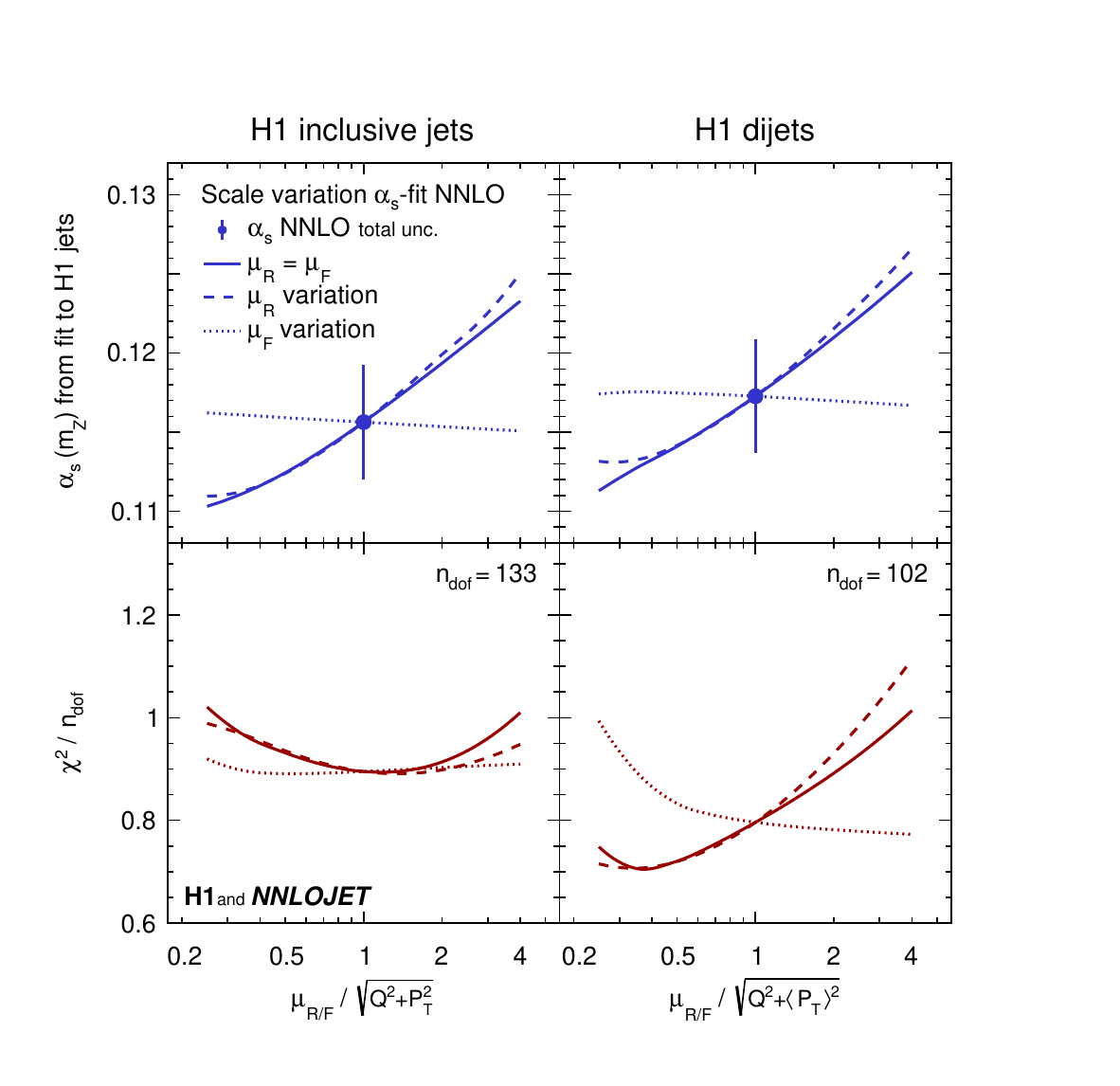}
\end{center}
\caption{
  Dependencies of the fitted values of \asmz\ as a function of the scale
  factors applied to the renormalisation and factorisation scales
  (\mur\ and \muf) for separate fits of inclusive jet and dijet data.
  The upper panels show the fitted value of \asmz, and the lower panels show the values of $\chisq/\ndf$.
  The left (right) panels show the values for the fit to inclusive jet
  (dijet) cross sections.
  The solid lines show the effects from varying \mur\ and \muf\ together.
  The dashed (dotted) lines show the effects from varying \mur\ (\muf) alone.
}
\label{fig:fit_scalevar}
\end{figure}

\begin{figure}[ht]
\begin{center}
   \includegraphics[width=0.48\textwidth]{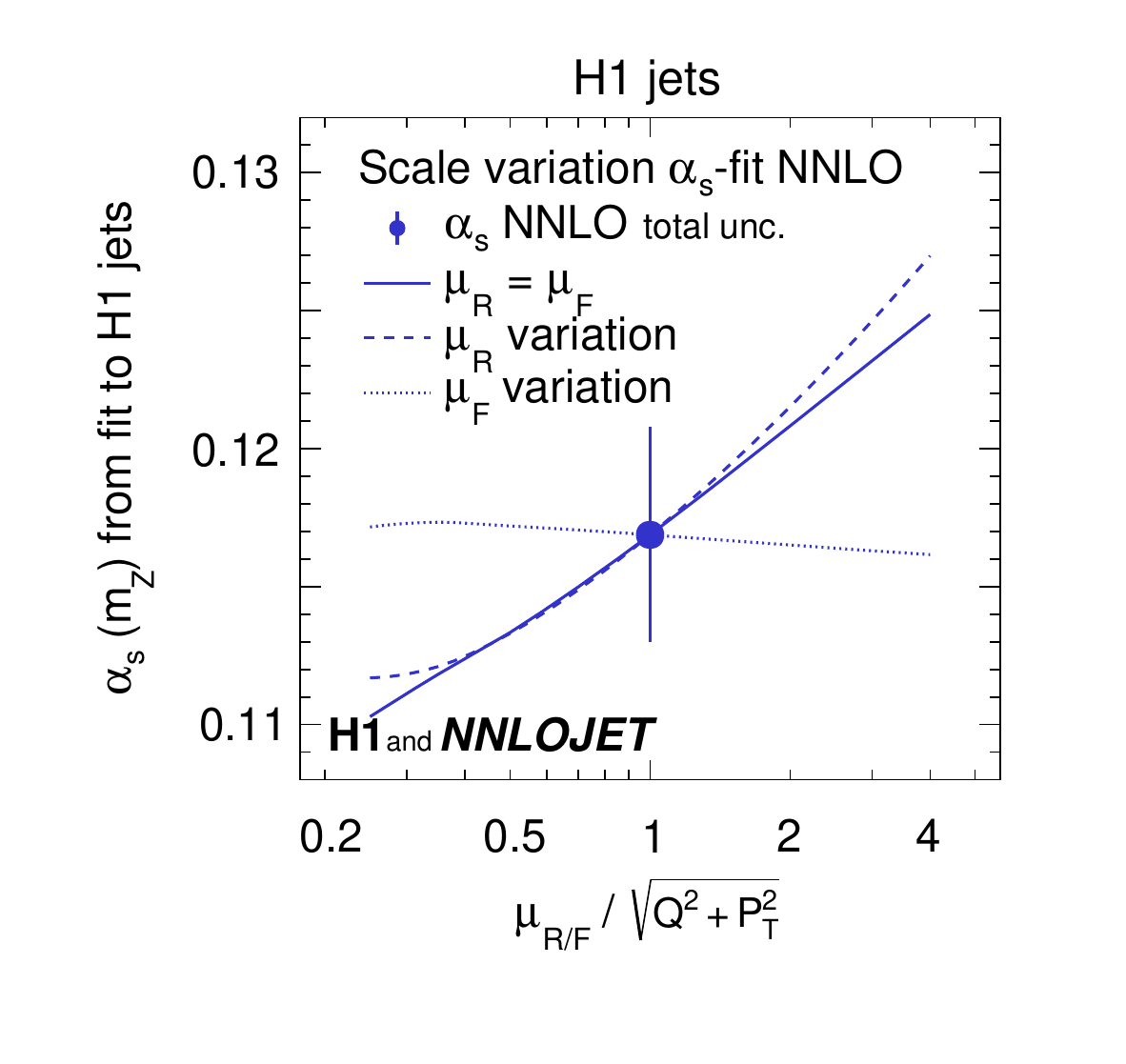}\hfill
   \includegraphics[width=0.48\textwidth]{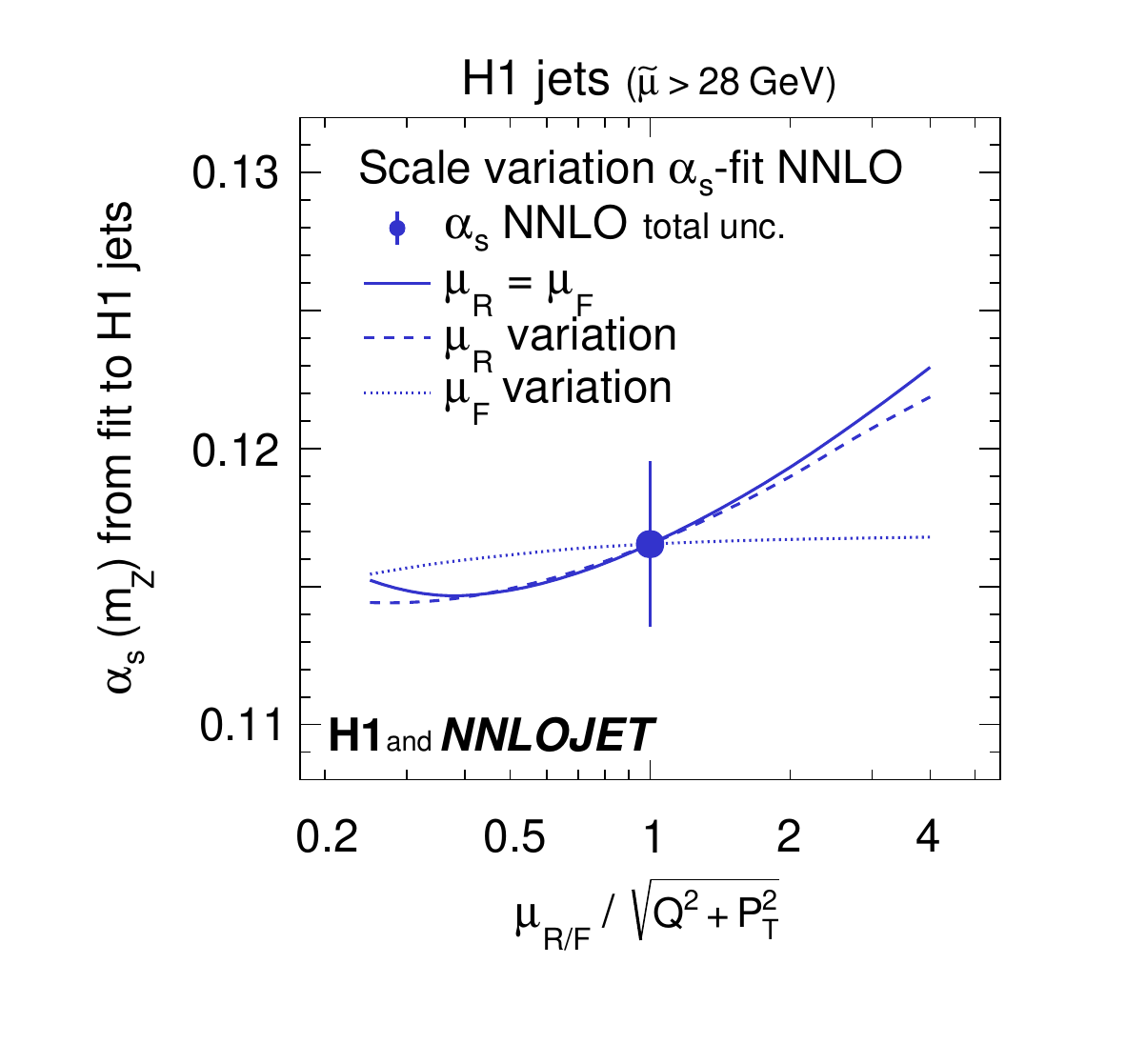}
\end{center}
\caption{
  Dependencies of the fitted values of \asmz\ as a function of the scale
  factors for the H1 jets fit (left) and the H1 jets fit with
  $\tilmu>28\,\GeV$ (right).
  Further details are given in the caption of figure~\ref{fig:fit_scalevar}.
}
\label{fig:fit_scalevar_MJ}
\end{figure}

The dependence of the results on \mur\ and \muf\
is studied by applying scale factors to the definition of
\mur\ and \muf. 
The values of \asmz\ and $\chisq/\ndf$ resulting from the fits  to
inclusive jet and to dijet cross sections are displayed in
figure~\ref{fig:fit_scalevar} %and for fits to H1 jets in figure~\ref{fig:fit_scalevar_MJ}
indicating that the standard choice for the scales (unity
scale factor) yields good values of $\chisq/\ndf$.
Figure~\ref{fig:fit_scalevar_MJ} displays the resulting \asmz\ for
fits to H1 jets.
In general, variations of \mur\ have a larger impact on the
result than those of \muf.
When restricting the data to $\tilmu>28\,\GeV$, the scale dependence
is greatly reduced.

Scale uncertainties are estimated through repeated fits with
scale factors applied simultaneously to \mur\ and \muf.
Instead of varying the scales up and down by conventional factors, in
this analysis a linear 
error propagation to the scale factors of 0.5 and 2 is performed
using the derivative determined at the nominal scale.
This is justified by the almost linear dependence on the logarithm of
the scale factor (figure~\ref{fig:fit_scalevar} and
\ref{fig:fit_scalevar_MJ}) and thus symmetric scale uncertainties are
presented.

\begin{figure}[ht]
\begin{center}
   \includegraphics[width=0.99\textwidth]{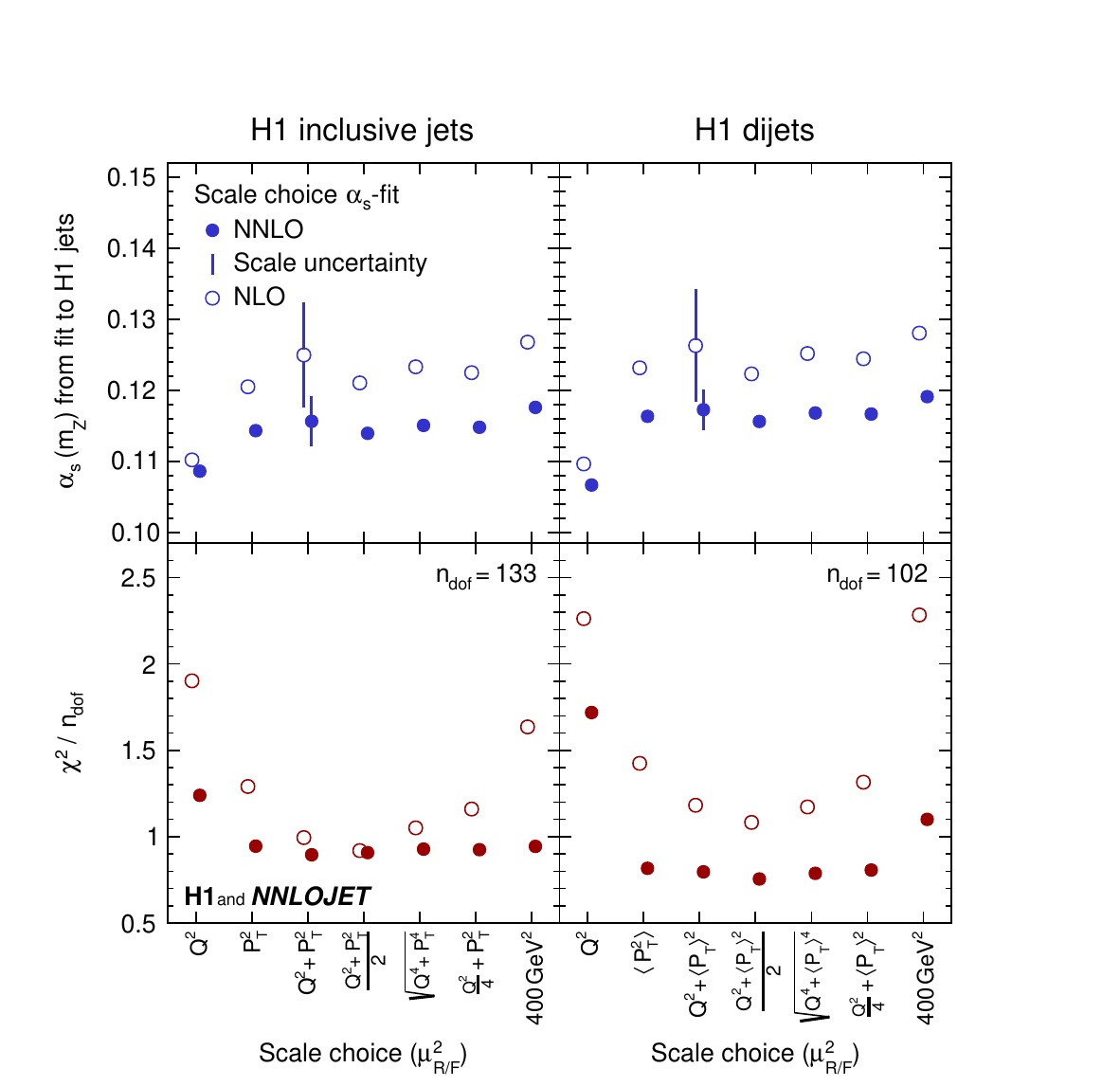}
\end{center}
\caption{
  Values of \asmz\ obtained for different 
  definitions of the renormalisation and factorisation scales
  (\mur\ and \muf)
  in separate fits of inclusive jet and dijet data.
  The lower panels show \chisq/\ndf\ of the fits.
  The open circles display results obtained using NLO matrix elements.
  The vertical bars indicate the scale uncertainties displayed
  together with the nominal scale choice.
}
\label{fig:fit_scalechoice}
\end{figure}

\begin{figure}[ht]
\begin{center}
   \includegraphics[width=0.48\textwidth]{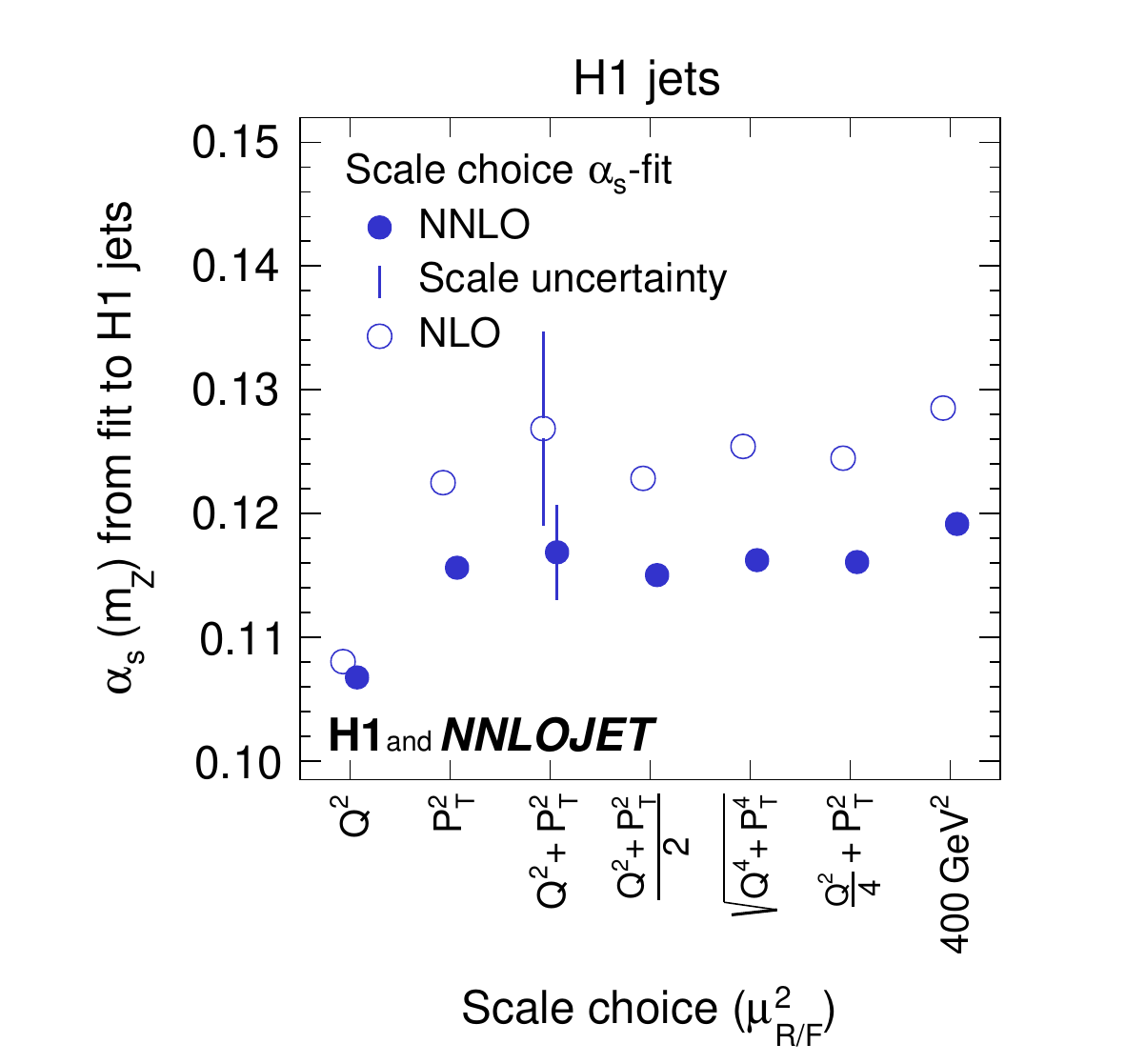}\hfill
   \includegraphics[width=0.48\textwidth]{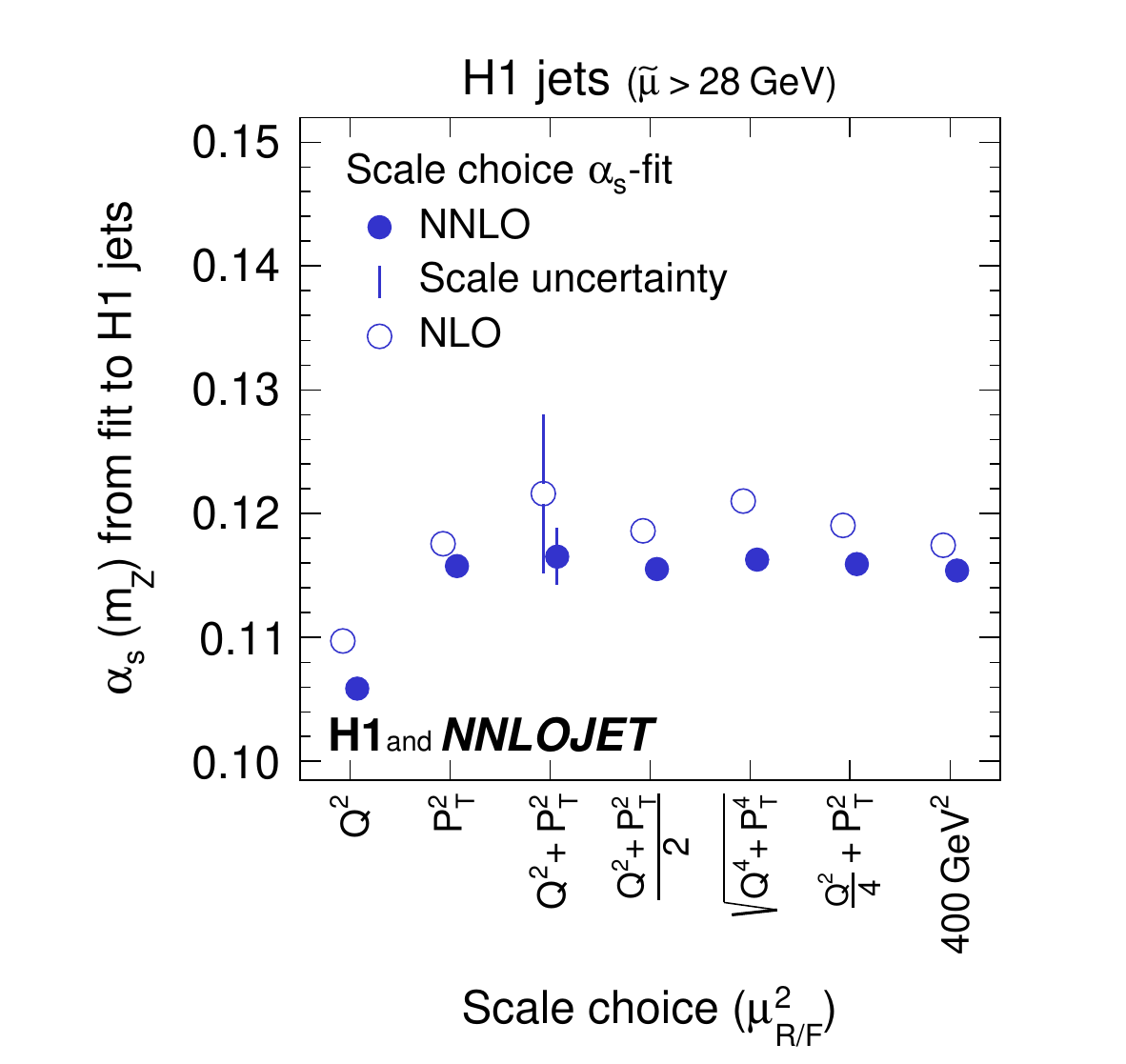}
\end{center}
\caption{
  Values of \asmz\ obtained for different 
  definitions of the renormalisation and factorisation scales
  (\mur\ and \muf)
  in the H1 jets fit (left) and the H1 jets fit with $\tilmu>28\,\GeV$
  (right).
  The open circles display results obtained using NLO matrix elements.
  The vertical bars indicate the scale uncertainties displayed
  together with the nominal scale choice.
  %Further details are given in the caption of
  %figure~\ref{fig:fit_scalechoice}.
}
\label{fig:fit_scalechoice_MJ}
\end{figure}

% --- scale choices and NLO
Alternative choices for \mur\ and \muf\ are investigated and the
results for \asmz\ with values of \chisq/\ndf\ are displayed in
figure~\ref{fig:fit_scalechoice} for fits to inclusive jet and dijet
data. 
The nominal scale definition $\mur^2=\muf^2=\Qsq+\pt^2$
results in good agreement of theory and data in terms of \chisq/\ndf.
The results obtained with alternative scale choices typically vary within the
assigned scale uncertainty. 
This is also observed for fits to H1 jets, presented in
figure~\ref{fig:fit_scalechoice_MJ}. 
A representative scale of the jet data analysed here is $20\,\GeV$.
Using $\mur=\muf=\mup=20\,\GeV$, simplifies the theory calculations such that
equations~\ref{eq:running}--\ref{eq:PDF} are not used and no
running of the coupling or evolution of the PDFs is needed.
For this scale choice the resulting value of \asmz\ (which after the fit is
evolved from $20\,\GeV$ to \mz\ for comparisons) is also found to be
consistent with the values obtained using the nominal scale.

The fits are repeated with the partonic cross sections $\hat{\sigma}_{i,k}$
calculated only up to NLO where for better comparisons identical
scale definitions and identical PDFs determined in NNLO fits are used.
For inclusive jets, the values of \chisq/\ndf\ of the NLO fits are of
comparable size for some of the studied scale choices, but are
significantly worse for certain choices such  
as $\mur^2=\muf^2=\Qsq$.
For dijets, the values of \chisq/\ndf\ are always higher for NLO than
for NNLO calculations.
The NLO calculations exhibit an enhanced
sensitivity to the choice of the scale and to scale
variations, as compared to NNLO, 
resulting in scale uncertainties of \asmz\ of $0.0077$,
$0.0081$ and $0.0083$ for inclusive jets, dijet and H1 jets, respectively,
as compared to uncertainties of $0.0034$, $0.0033$ and $0.0038$ in
NNLO, respectively. 
The previously observed reduction of scale uncertainties of the cross
section predictions at
NNLO~\cite{Currie:2016ytq,Andreev:2016tgi,Currie:2017tpe} is reflected
in a corresponding reduction of the \asmz\ scale uncertainties.

\begin{figure}[ht]
\begin{center}
   \includegraphics[width=0.7\textwidth]{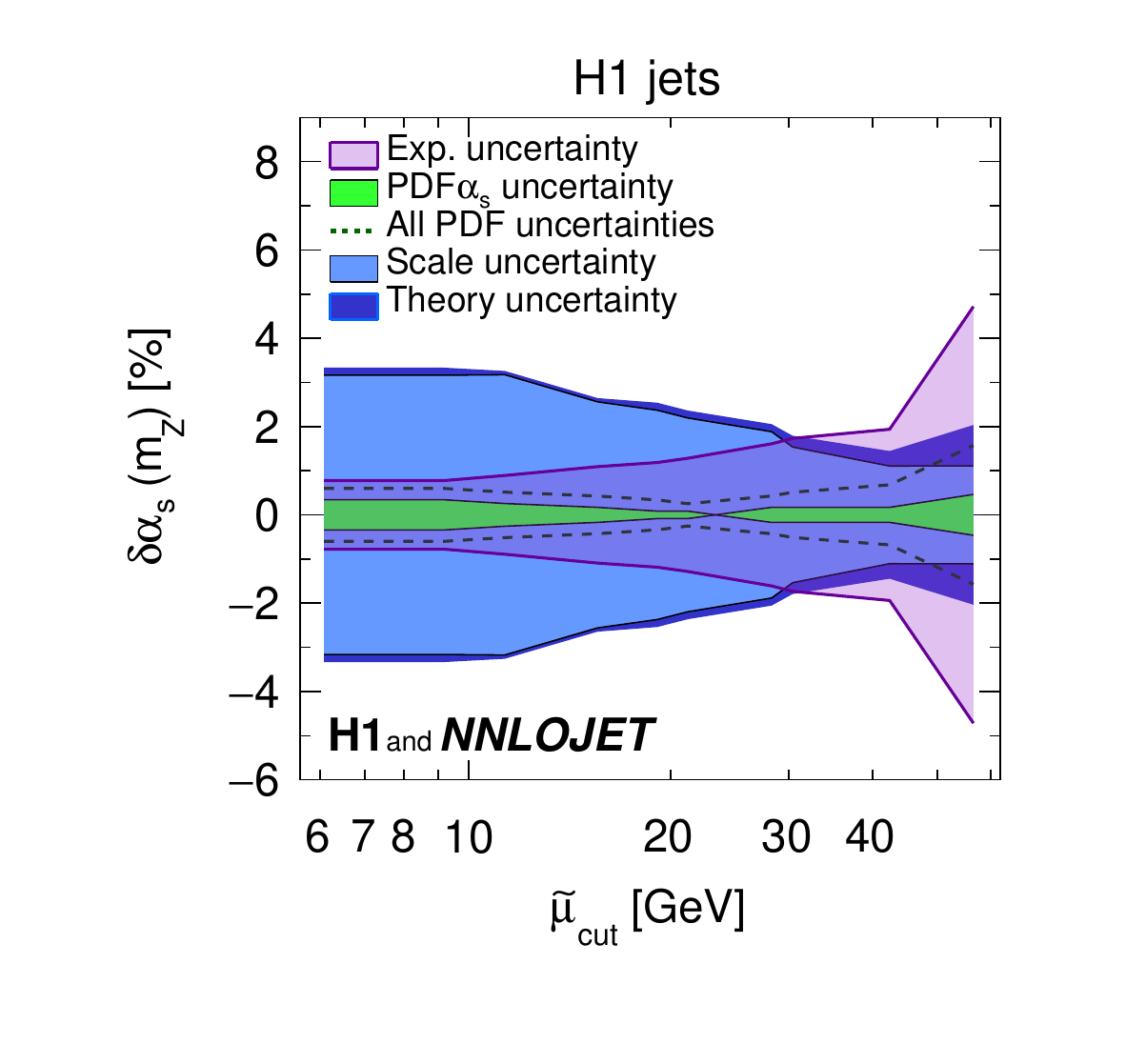}
\end{center}
\caption{
  Uncertainties of the \as\ fit as a function of the
  parameter $\tilmu_{\rm cut}$ which restricts the jet data to high scales.
  The experimental, scale, PDF\as, quadratic sum of all PDF related
  uncertainties, and the theory uncertainty are shown.
}
\label{fig:uncertainties}
\end{figure}

%%%%%%%%%%%%%%%%%%%%%%%%%%%%%%%%%%%%%%%%%%%%%%%%%%%%%%%%%%%%
% mu-cut
\paragraph{Restricting the scale \boldmath\tilmu}
\label{sec:mucut}
% ---- scale/running dependence of uncertainties
In order to study the size of the uncertainties
as a function of $\tilmu$, 
the fits to inclusive jet and to dijet cross
sections are repeated using data points exceeding a given
value $\tilmu_{\rm cut}$. %, which is set to $\m_b$ for our fits.
The resulting uncertainties are displayed in figure~\ref{fig:uncertainties}.  
The experimental uncertainties are smaller for lower $\tilmu$.
This is because more data are considered in the fit, but also since
the data at lower values of $\tilmu$ have an enhanced
sensitivity to \asmz\ due to the running of the strong coupling.
In contrast, the scale uncertainties of the NNLO cross section
predictions are largest for low values of $\tilmu$, and thus decrease
with increasing $\tilmu$. 
Considering only data with values of $\tilmu$ above approximately
$30\,\GeV$ the experimental and scale
uncertainty become similar in size.

The result obtained with $\tilmu>28\,\GeV$ is considered as the main
result of this article.

At values of $\tilmu_{\rm cut}$ around $20\,\GeV$ the PDF\as\ uncertainty
effectively vanishes.
In other words, the fit result is insensitive to the
\asmzPDF\ assumptions made for the PDF determination.
A possible explanation is the gradual change of the
fraction of gluon and quark induced processes with \tilmu: data at
lower values of \tilmu\ have contributions from low-$x$ where the
gluon PDF is dominating,  whereas data at higher values of
\tilmu\ have a successively higher fraction of quark induced processes.
The quark PDFs are less dependent on \asmzPDF\ than the gluon PDF, 
and are well determined by inclusive DIS data.

%%%%%%%%%%%%%%%%%%%%%%%%%%%%%%%%%%%%%%%%%%%%%%%%%%%%%%%%%%%%
%                      Results
%%%%%%%%%%%%%%%%%%%%%%%%%%%%%%%%%%%%%%%%%%%%%%%%%%%%%%%%%%%%
%\clearpage
\subsection{Results}
\begin{figure}[ht]
\begin{center}
   \includegraphics[width=0.7\textwidth]{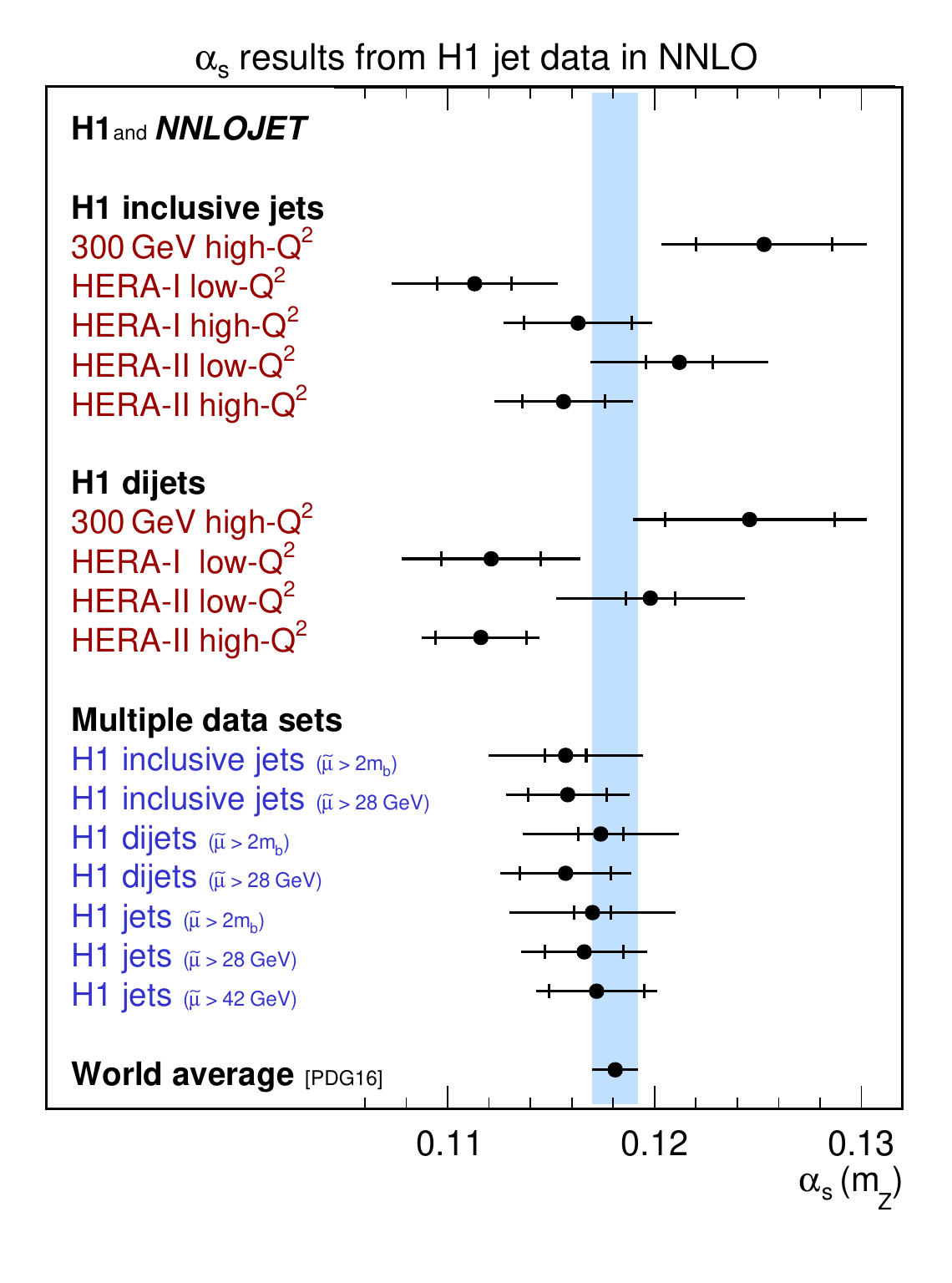}
\end{center}
\caption{
  Summary of \asmz\ values obtained from fits to individual
  and multiple H1 jet data sets. The inner error bars
  indicate the experimental uncertainty and the outer error bars the
  total uncertainty. 
}
\label{fig:summary}
\end{figure}

%%%%%%%%%%%%%%%%%%%%%%%%%%%%%%%%%%%%%%%%%%%%%%%%%%%%%%%%%%%%
%   main results
\paragraph{The value of the strong coupling constant \boldmath\asmz}
\label{sec:results}
The values of \asmz\ obtained from the fits to the data are
collected in table~\ref{tab:asresults} and displayed in
figure~\ref{fig:summary}. 
Good agreement between theory and data is found. 

% individual data sets
For the fits to the individual data sets the \chisq/\ndf\ is below
unity in most cases.
The \asmz\ values are all found to be consistent, in
particular between inclusive jet and dijet measurements.

% inclusive jets
The fits to the inclusive jet data exhibit very reasonable \chisq/\ndf\ values,
thus indicating the consistency of the individual data sets.
The value of \asmz\ from `H1 inclusive jets' has a significantly
reduced experimental uncertainty compared to the results for the
individual data sets.
The cut $\tilmu>28\,\GeV$ results for inclusive jets in  
$\asmz=0.1158\,(19)_{\rm exp}\,(23)_{\rm th}$,
which is consistent with the world average~\cite{Bethke:2017uli,Olive:2016xmw}. 

% dijets
Value of \chisq/\ndf\ lower than unity are obtained for fits to all dijet cross sections
confirming their consistency.
The results agree with those from inclusive jet cross sections and the
world average.
At high scales $\tilmu>28\,\GeV$, a value $\asmz=0.1157\,(22)_{\rm exp}\,(23)_{\rm th}$ is found.

% H1 jets
The fit to H1 jets yields $\chisq/\ndf = 0.87$ for 200
data points and $\asmz=0.1170\,(9)_{\rm  exp}\,(38)_{\rm th}$. 
The scale uncertainty is the largest among the theoretical
uncertainties and all other uncertainties are negligible in comparison.

The \asmz\ value obtained from H1 jet data restricted to $\tilmu>28\,\GeV$ is
\begin{align}
  \asmz = 0.1166\,(19)_{\rm exp}\,(9)_{\rm had}\,(3)_{\rm PDF}\,(2)_{\rm PDF\as}\,(4)_{\rm PDFset}\,(21)_{\rm scale}
\nonumber
\end{align} 
with $\chisq= 62.4$ for 91 data points.
Although the reduced number of data points leads to an increased
experimental uncertainty, as compared to the option 
$\tilmu>2m_b$, it is still smaller than the scale uncertainty,
which is found to be reduced significantly. 
All PDF related uncertainties essentially vanish\footnote{
  The difference of the
  main fit result to $\asmzPDF=0.118$
  is covered by the  systematic variation $\asmzPDF=0.118\pm0.002$.
}.
Therefore, this \asmz\ determination is taken as the main result.
This result as well as those results obtained from the inclusive jet
and dijet data separately are consistent with the world average.

The main result is also found to be consistent with 
$\asmz=0.1165(8)_{\text{exp}}(38)_{\text{pdf,theo}}$ determined previously in
NLO accuracy from normalised H1 HERA-II
high-\Qsq\ jet cross section data~\cite{Andreev:2014wwa}.
That result is experimentally more precise, mainly because data at
somewhat lower scales and three-jet data are
included\footnote{No NNLO calculation is available for
  three-jet production in DIS to date}. 
The scale uncertainty of the previous NLO fit is larger than for 
the present analysis in NNLO, despite of the fact that it was considered to be
partially uncorrelated bin-to-bin in the previous NLO fit, whereas the present
approach is more conservative.

In the present analysis, the value with the smallest total
uncertainty is obtained in a fit to H1 jets restricted to
$\tilmu>42\,\GeV$ with the result $\asmz=0.1172\,(23)_{\rm
  exp}\,(18)_{\rm theo}$  and a value of $\chisq/\ndf=37.0/40$.
This result, however, is obtained from a very limited number of
measurements, the precision of which is limited by statistical uncertainties.

\begin{figure}[ht]
\begin{center}
  \includegraphics[width=0.8\textwidth]{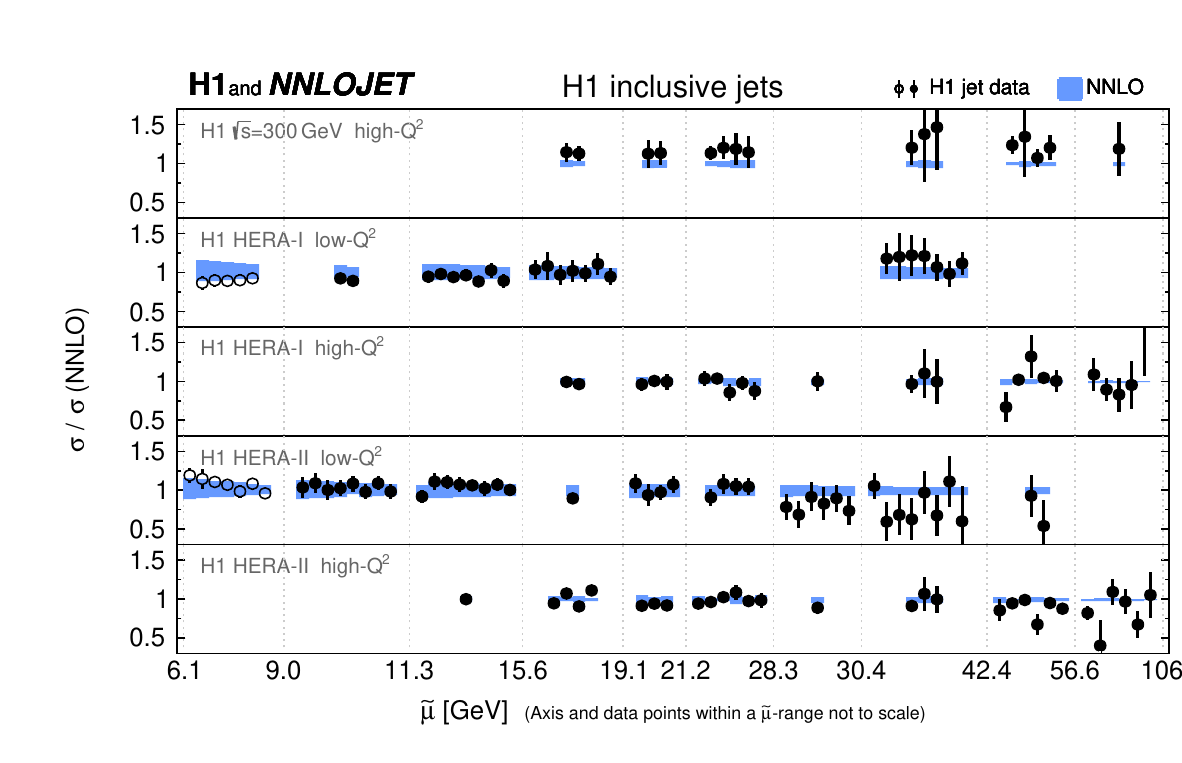}
  \includegraphics[width=0.8\textwidth]{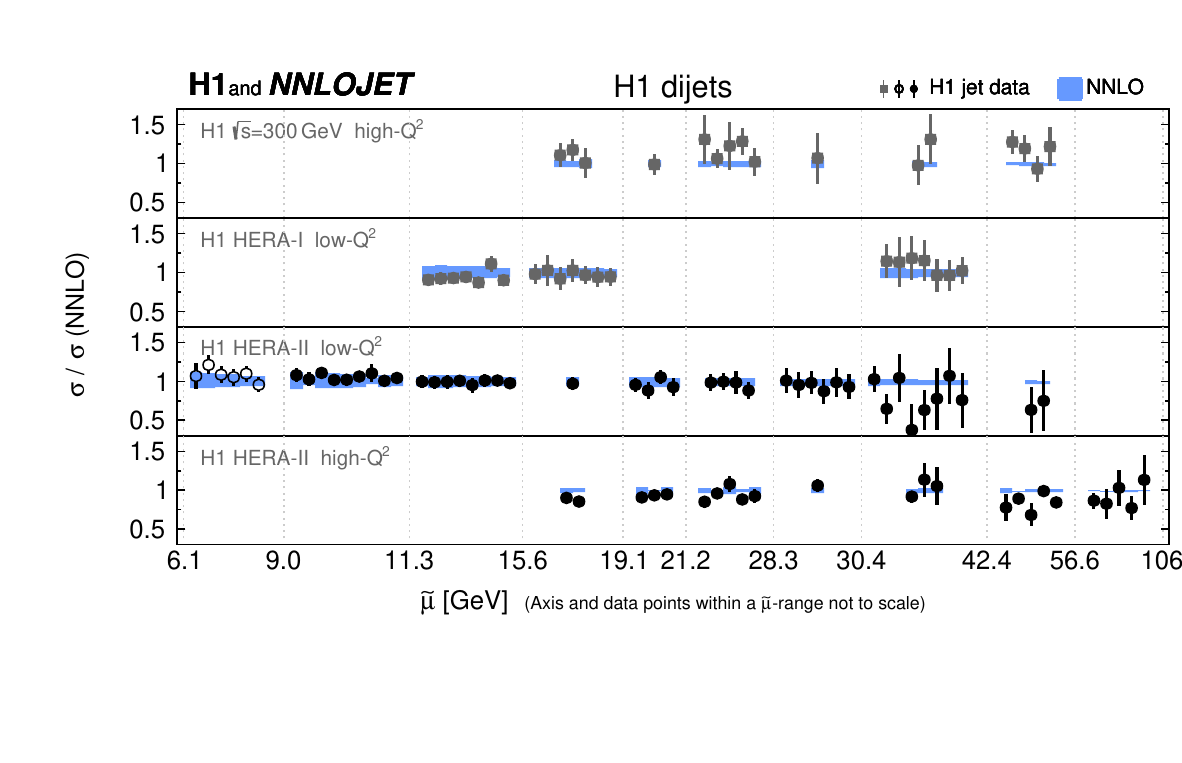}
\end{center}
\caption{
  Ratio of inclusive jet (upper panel) and dijet cross sections (lower
  panel) to NNLO predictions obtained with the fitted value $\asmz=0.1157$.
  Data points are ordered according to their scale
  \tilmu\ and are displayed on the horizontal axis within the
  respective \tilmu-interval. Within a single interval
  multiple data points are displayed with equal horizontal spacing
  and are thus not to scale. 
  The displayed intervals reflect the choices made for the
  studies of the running of the strong coupling (compare
  figures~\ref{fig:running_H1} and~\ref{fig:running_MJ}).
  The shaded area indicates the uncertainty on the NNLO calculations
  from scale variations. The open circles show data points which are
  not considered for some fits, because their scale $\tilmu$ is
  below  $2m_b$.
  The squares show data points not considered for the `H1
  jets'-fit, since the statistical correlations to the respective
  inclusive jet measurements are not known. 
}
\label{fig:dataratio}
\end{figure}

% data to NNLO plots
The ratio of all H1 jet cross section measurements to the NNLO predictions
is displayed in figure~\ref{fig:dataratio}.
Overall good agreement between data and predictions is observed.

%%%%%%%%%%%%%%%%%%%%%%%%%%%%%%%%%%%%%%%%%%%%%%%%%%%%%%%%%%%%
% --- running
\paragraph{Running of the strong coupling constant}
\label{sec:running}
The strong coupling is determined in fits to data points grouped into
intervals $[\tilmu_{\rm lo};\tilmu_{\rm up}]$ of \tilmu.
The data point grouping and the interval boundaries can be read off
figure~\ref{fig:dataratio}.
The assumptions on the running of \asmur\ thus are for each fit restricted
to a limited \mur\ range\footnote{For purely technical reasons the fit parameter is \asmz, and thus the
running is applied from \mur, as used in the calculation, to \mz\ and
then `back' to a representative value \mur{}.}.
For a given data point its \tilmu\ value is representative for the
\mur\ range probed by the corresponding prediction, see equations \ref{eq:geommean} and \ref{eq:scaledefinition}.
The fit results are for each interval shown at the representative scale $\mur=\sqrt{\tilmu_{\rm lo}\tilmu_{\rm up}}$.

\begin{figure}[ht]
\begin{center}
   \includegraphics[width=0.7\textwidth]{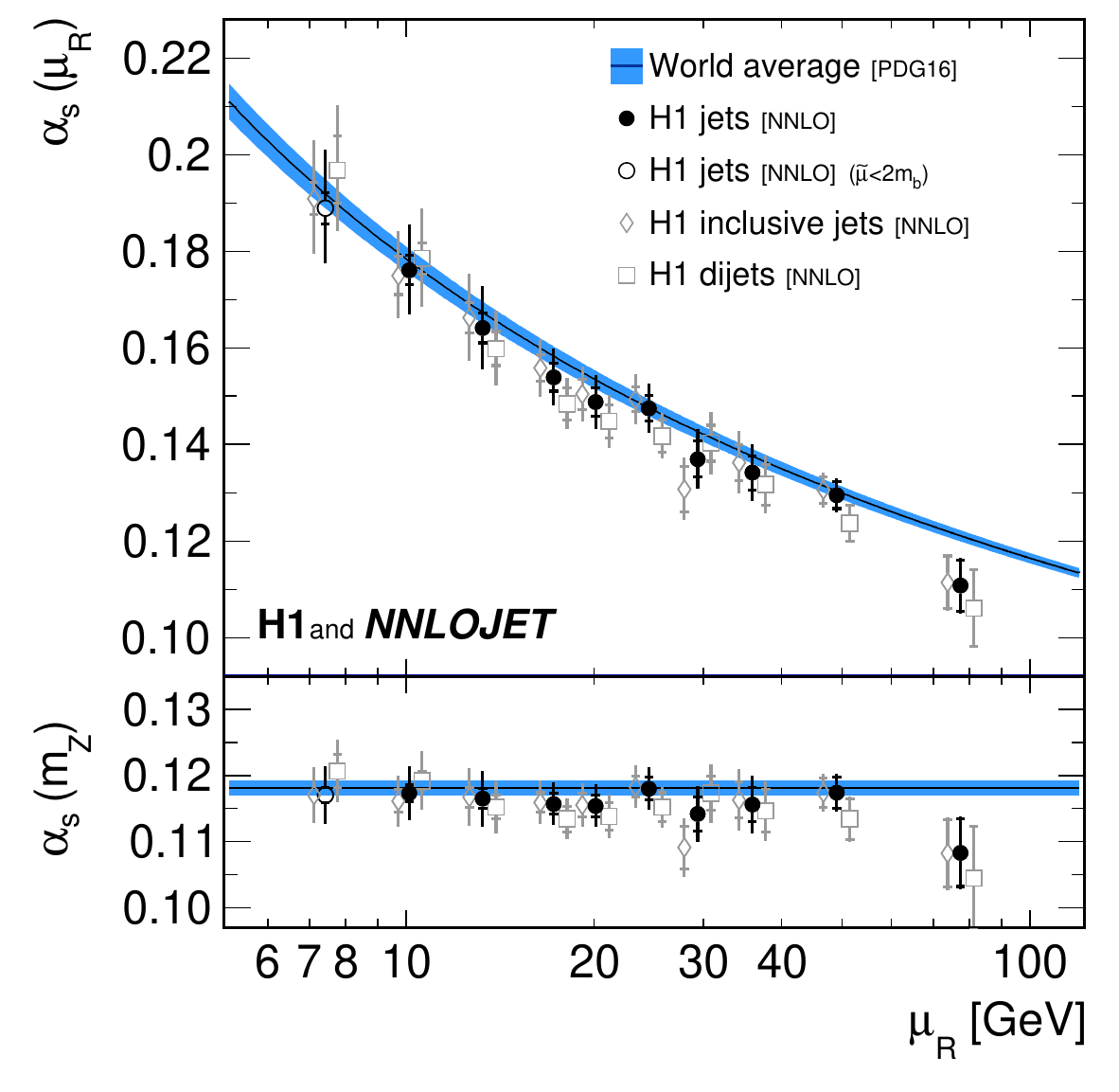}
\end{center}
\caption{
  Results on \asmz\ and \asmur\ for fits to data points arranged in groups of similar
  \mur. The circles show results from 
  inclusive jet and dijet data taken together (`H1 jets'), the open
  diamonds results from inclusive jet cross sections alone and the open boxes
  results from dijet cross sections alone.
  For these fits, the data sets are not constrained by the requirement
  $\tilmu>2m_b$. 
  The fitted values of \asmz\ (lower panel) are translated to
  \asmur\ (upper panel), using the
  solution of the QCD renormalisation group equation. 
  The data points from fits to inclusive jets (dijets) are
  displaced to the left (right) for better visibility.
  In the upper panel a displacement is also applied along the vertical direction, to account for the running of \asmur.
  The inner error bars denote the experimental uncertainties alone,
  and the outer error bars indicate the total  uncertainties.  
}
\label{fig:running_H1}
\end{figure}

The results for fits to inclusive jet and to dijet cross sections, as
well as to H1 jets, are presented for the ten selected intervals in
$\tilmu$ in table~\ref{tab:running} and are
displayed in figure~\ref{fig:running_H1}.
Consistency is found for the fits to inclusive jets, dijets, and H1
jets, and the running of the strong  coupling is confirmed in the
accessible range of approximately 7 to $90\,\GeV$.
The lowest interval considered contains the data points with
$\tilmu<2m_b$, which are excluded from the main analysis.
Nevertheless, these results are found to be consistent with the
other \asmz\ determinations presented here.

\begin{figure}[ht]
\begin{center}
   \includegraphics[width=0.7\textwidth]{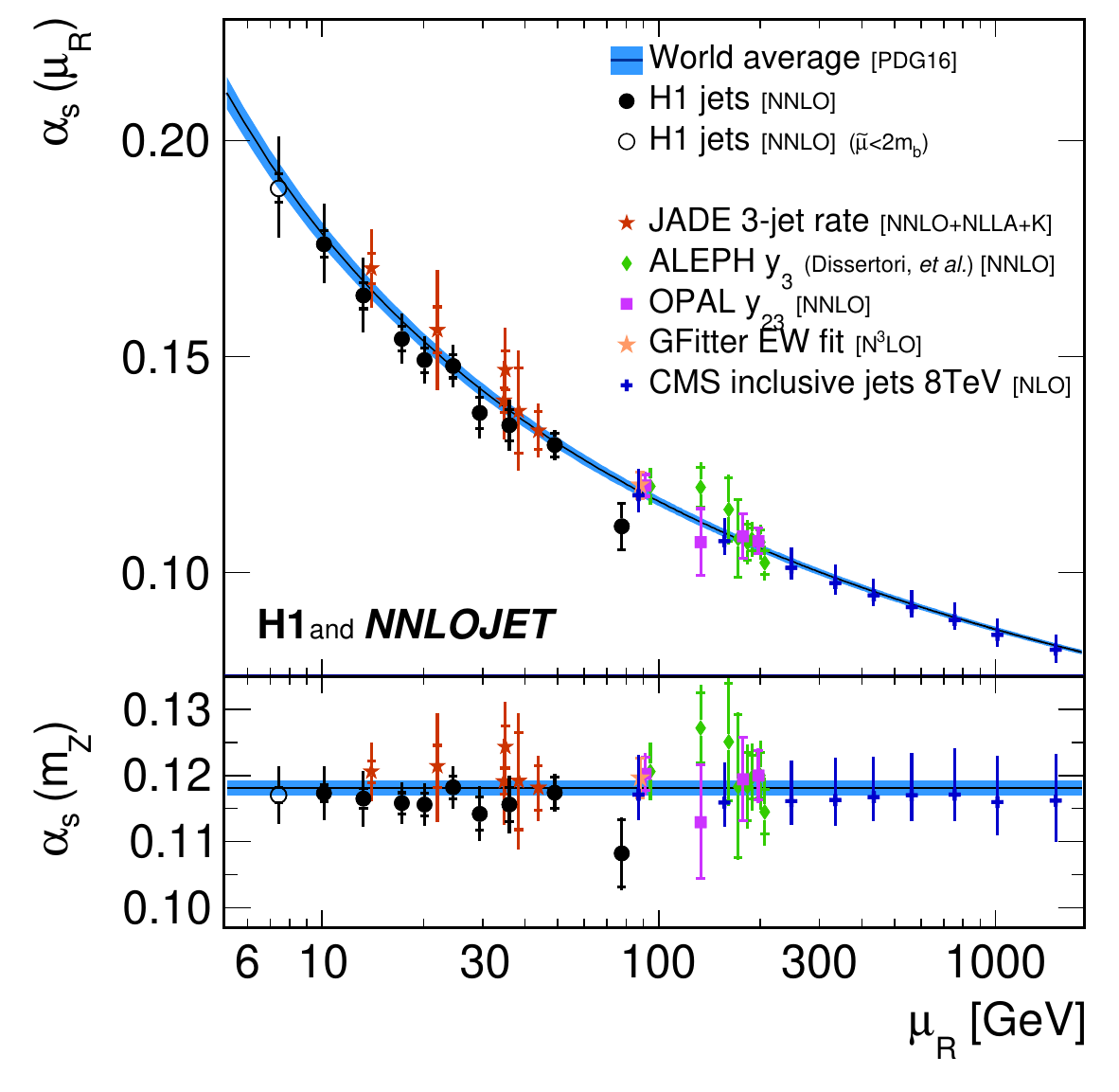}
\end{center}
\caption{
  Results for \asmz\ and \asmur\ for fits to data points arranged in groups of
  similar \mur, compared to results from other experiments and
  processes.
  Further details can be found in the caption of figure~\ref{fig:running_H1}.
}
\label{fig:running_MJ}
\end{figure}

The values obtained from fits to H1 jets are compared to other
determinations of at least NNLO
accuracy~\cite{Baak:2014ora,OPAL:2011aa,Schieck:2012mp,Dissertori:2007xa} %,Chatrchyan:2013haa}
and to results at NLO at very high scale~\cite{Khachatryan:2016mlc}
in figure~\ref{fig:running_MJ}, and consistency with the other experiments 
is found.

The results are consistent with results obtained from an alternative
method used as a cross check, where in a single fit with ten free
parameters the \as\ values in the ten bins are determined
simultaneously.

%%%%%%%%%%%%%%%%%%%%%%%%%%%%%%%%%%%%%%%%%%%%%%%%%%%%%%%%%%%%
%                    PDF+as fit
%%%%%%%%%%%%%%%%%%%%%%%%%%%%%%%%%%%%%%%%%%%%%%%%%%%%%%%%%%%%
%\clearpage
\begin{boldmath}
\section{Simultaneous \as\ and PDF determination}
\label{sec:pdfasfit}
\end{boldmath}
In addition to the fits described above also a fit in NNLO accuracy of
\asmz\ together with the non-perturbative PDFs is performed which
takes jet data and inclusive DIS data as input.
This fit is denoted as `PDF+\as-fit' in the following.

\subsection{Methodology}
The methodology of the PDF+\as-fit is closely related to PDF
determinations as performed by other
groups~\cite{Alekhin:2017kpj,Dulat:2015mca,Abramowicz:2015mha,Harland-Lang:2014zoa,Ball:2017nwa}.
The PDFs are parametrised at a low starting scale $\mu_0$ which is
below the charm-quark mass. Heavy-quark PDFs are 
generated dynamically and only light-quark PDFs and the gluon
distribution have to be determined in the fit.

% --- inclusive DIS data
In order to have constraints on the PDFs, polarised and
unpolarised inclusive NC and CC DIS cross
sections~\cite{Adloff:1999ah,Adloff:2000qj,Adloff:2003uh,Collaboration:2010ry,Aaron:2012qi,Aaron:2012kn}
are used (table~\ref{tab:DISdata}).
This data sample is identical to the one used in the H1PDF2012 PDF
fit~\cite{Aaron:2012qi}.
%
% --- normalised jet cross sections
In addition, normalised inclusive jet and dijet cross
sections~\cite{Aktas:2007aa,Andreev:2014wwa,Andreev:2016tgi} are used
(table~\ref{tab:normdata}).

% --- predictions
The calculations of the splitting kernels are performed in NNLO
using the program QCDNUM~\cite{Botje:2010ay,Botje:2016wbq}.
The predictions for the inclusive DIS cross sections are calculated using
structure function calculations in NNLO using the zero-mass
variable flavour number scheme (ZM-VFNS)~\cite{Aaron:2012qi} as implemented in
QCDNUM~\cite{Botje:2010ay,Botje:2016wbq}.
Normalised jet cross sections are calculated as a ratio of
jet cross sections to inclusive NC DIS, where the former are calculated
as outlined in section~\ref{sec:predictions} and the latter are
calculated using ZM-VFNS structure functions using QCDNUM.
For inclusive DIS predictions the scales $\mur^2$ and $\muf^2$ are both set to \Qsq\
and for jet predictions to $\Qsq+\pt^2$, as specified in equation~\ref{eq:scale}.

% --- Q2 cut
For the PDF+\as-fit all data are restricted to the range
$\Qsq>10\,\GeVsq$ in order to exclude kinematic regions where
fixed-order pQCD
cannot be applied reliably.
For jet cross sections $\tilmu>2m_b$ is required in addition.
After applying these cuts, the jet predictions receive contributions from
the $x$-range down to $0.003$, whereas without these cuts it would be 0.002.
Major contributions to the data points at highest values of \tilmu\ are
within the $x$-range 0.1 to 0.5.

% --- parameterisation
The choice of the PDF parametrisations and the values of input
parameters follows closely previous
approaches~\cite{Adloff:2003uh,Abramowicz:2015mha,Aaron:2012qi,Klein:1564929}
and are only discussed briefly here.
At a starting scale $\mu_0^2=1.9\,\GeVsq$ parton densities are attributed to the constituents of the proton.
These take the functional form
\begin{equation}
  xf(x)|_{\mup} = f_Ax^{f_B}(1-x)^{f_C}(1+f_Dx+f_Ex^2),
\label{eq:PDFparam}
\end{equation}
where $f$ is one of $g$, $\tilde{u}$, $\tilde{d}$, $\bar{U}$,
$\bar{D}$, denoting the density of the gluon, up-valence,
down-valence, up-sea, down-sea in the proton, respectively.
The strange sea is set to $\bar{s}(x)=f_s\bar{D}$, where $f_s=0.4$.
Parameters $f_D$ and $f_E$ are set to zero by default, but are added
for specific flavours in order to improve the fit. The parameters
$g_A$, $\tilde{u}_A$ and $\tilde{d}_A$ are constrained by sum
rules. The parameter $\bar{U}_A$ is set equal to
$\bar{D}_A(1-f_s)$. The parameter $\bar{U}_B$ is set equal to
$\bar{D}_B$. A total of $12$ fit parameters are used to describe the PDFs.

% --- error estimation
The uncertainty obtained from the fit comprises experimental
uncertainties of the data and hadronisation uncertainties of the jet
cross section predictions.
The resulting uncertainty of \asmz\ from the PDF+\as-fit is denoted as `exp,had,PDF'.
In order to determine also model (`mod') and parametrisation (`par')
uncertainties, an additional error estimation similar to HERAPDF2.0~\cite{Abramowicz:2015mha} is performed.
The model uncertainty is estimated as the quadratic sum of the differences
of the nominal result to the resulting values of \as\ when repeating the
PDF+\as-fit with alternative parameters, such as the charm or beauty masses or the sea quark
suppression factor $f_{\rm s}$~\cite{Abramowicz:2015mha}. 
Parametrisation uncertainties are attributed by adding extra $f_D$ or
$f_E$ parameters to the fit or by varying the starting scale.
In addition, a more flexible functional form is allowed for the gluon,
similar to the PDF parametrisation used for the default
HERAPDF2.0~\cite{Abramowicz:2015mha} fit\footnote{The functional form
  referred to as ``alternative gluon'' in \cite{Abramowicz:2015mha}
  actually corresponds to the default choice for this paper.}.
A total of eight parametric forms different from the default are considered.

The scale uncertainty of \asmz\ from this fit is determined by
repeating fits with scale
factors $0.5$ and $2$ applied to \mur\ and \muf\ simultaneously to all
calculations involved.
The larger of the two deviations from the central fit, corresponding
to a scale factor of $0.5$, is taken as
symmetric scale uncertainty.
A more detailed study is beyond the scope of this paper.

The PDF+\as-fit differs from the \as-fit
outlined in section~\ref{sec:asfit} in the following aspects:
the usage of normalised jet cross sections, the inclusion of NC and CC
DIS cross sections and the low 
starting scale $\mu_0$ of the DGLAP evolution, thus assuming the validity of
the running coupling and the PDF evolution down to lower scale values.

%   PDF+as fit result
\subsection{Results}
\label{sec:ResultsPDFasFit}
\paragraph{Fit results and the value of \boldmath\asmz}
The results of the PDF+\as-fit are presented in table~\ref{tab:PDFfit}.
The fit yields $\chisq/\ndf=1518.6/(1529-13)$, confirming
good agreement between the predictions and the data. 
The resulting PDF is able to describe 141 jet data points and the
inclusive DIS data simultaneously. 

\begin{figure}[ht]
\begin{center}
  \includegraphics[width=0.7\textwidth]{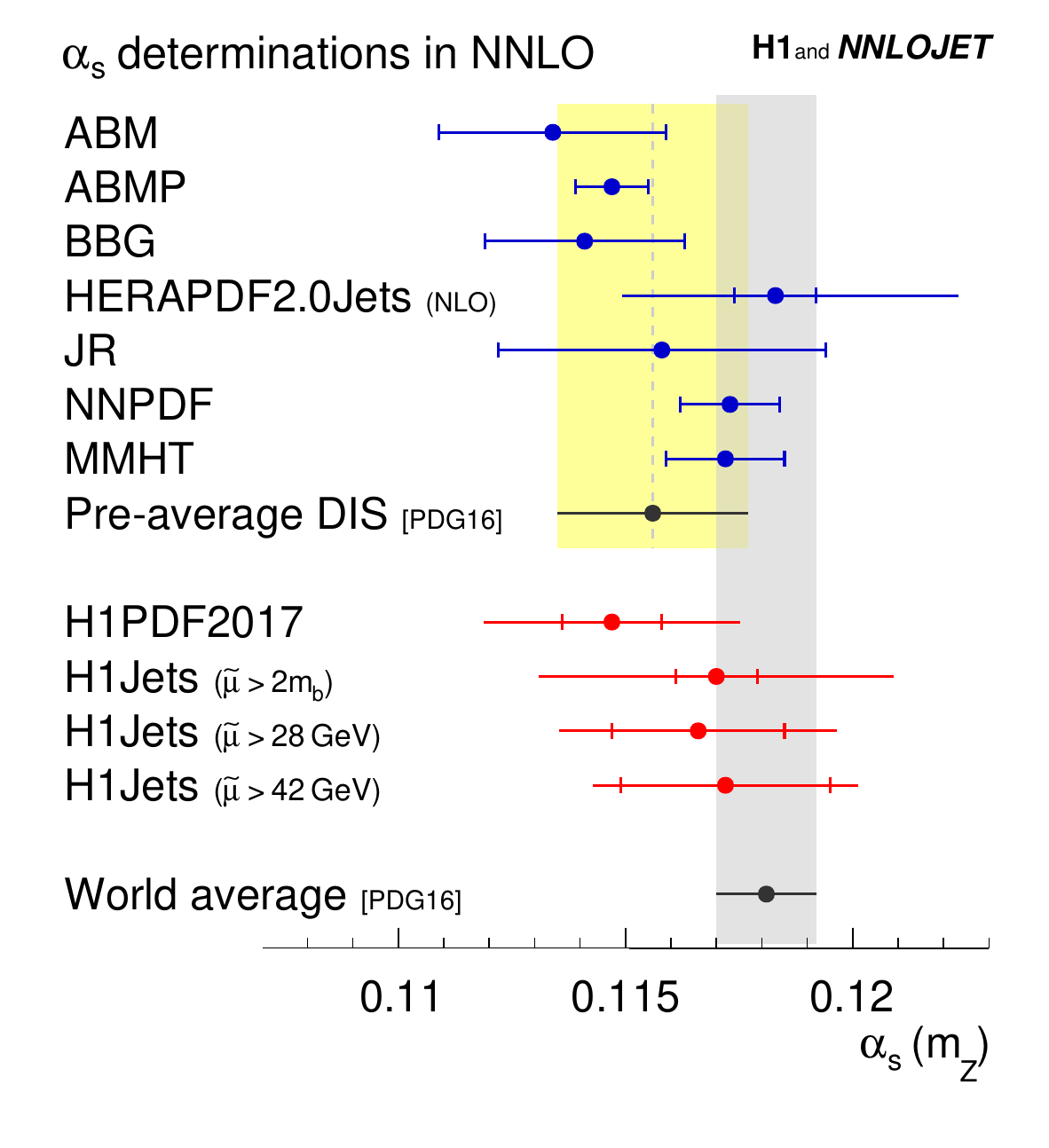}
\end{center}
\caption{
  Comparison of the value of \asmz\ obtained in the \HonePDF\ PDF+\as-fit and in the H1 jets fit in NNLO accuracy to other
  \as\ determinations from DIS data. 
  The pre-average of structure function data and the world
  average~\cite{Olive:2016xmw}  are also indicated.
}
\label{fig:pdffits}
\end{figure}

The value of \asmz\ is determined to
\begin{equation}
  \asmz=\PDFasResult\,(11)_{\rm exp,had,PDF}\,(2)_{\rm mod}\,(3)_{\rm par}\,(23)_{\rm scale}\,.
\nonumber
\end{equation}
and is determined to an overall precision of 2.2\,\%.
It is worth noting that the result is largely insensitive to
the PDF model and parametrisation choices. The scale uncertainty is
dominating.
The \asmz\ value is consistent with the main result of the `H1 jets'
fit.
The result is compared to values from the 
PDF fitting groups ABM~\cite{Alekhin:2012ig},
ABMP~\cite{Alekhin:2017kpj}, BBG~\cite{Blumlein:2006be},
HERAPDF~\cite{Abramowicz:2015mha}, JR~\cite{JimenezDelgado:2008hf},
NNPDF~\cite{Ball:2011us} and MMHT~\cite{Harland-Lang:2014zoa} in figure~\ref{fig:pdffits}
and consistency is found.
The value is consistent with the world average 
and the `pre-average' value of the structure function category~\cite{Olive:2016xmw}. 
The result exhibits a competitive experimental uncertainty to other
determinations~\cite{Ball:2011us,Harland-Lang:2014zoa,Alekhin:2017kpj},
which is achieved by using H1 normalised jet cross sections
in addition to the H1 inclusive DIS data. 

\begin{figure}[ht]
\begin{center}
  \hspace{-0.08\textwidth}
  \includegraphics[width=0.56\textwidth]{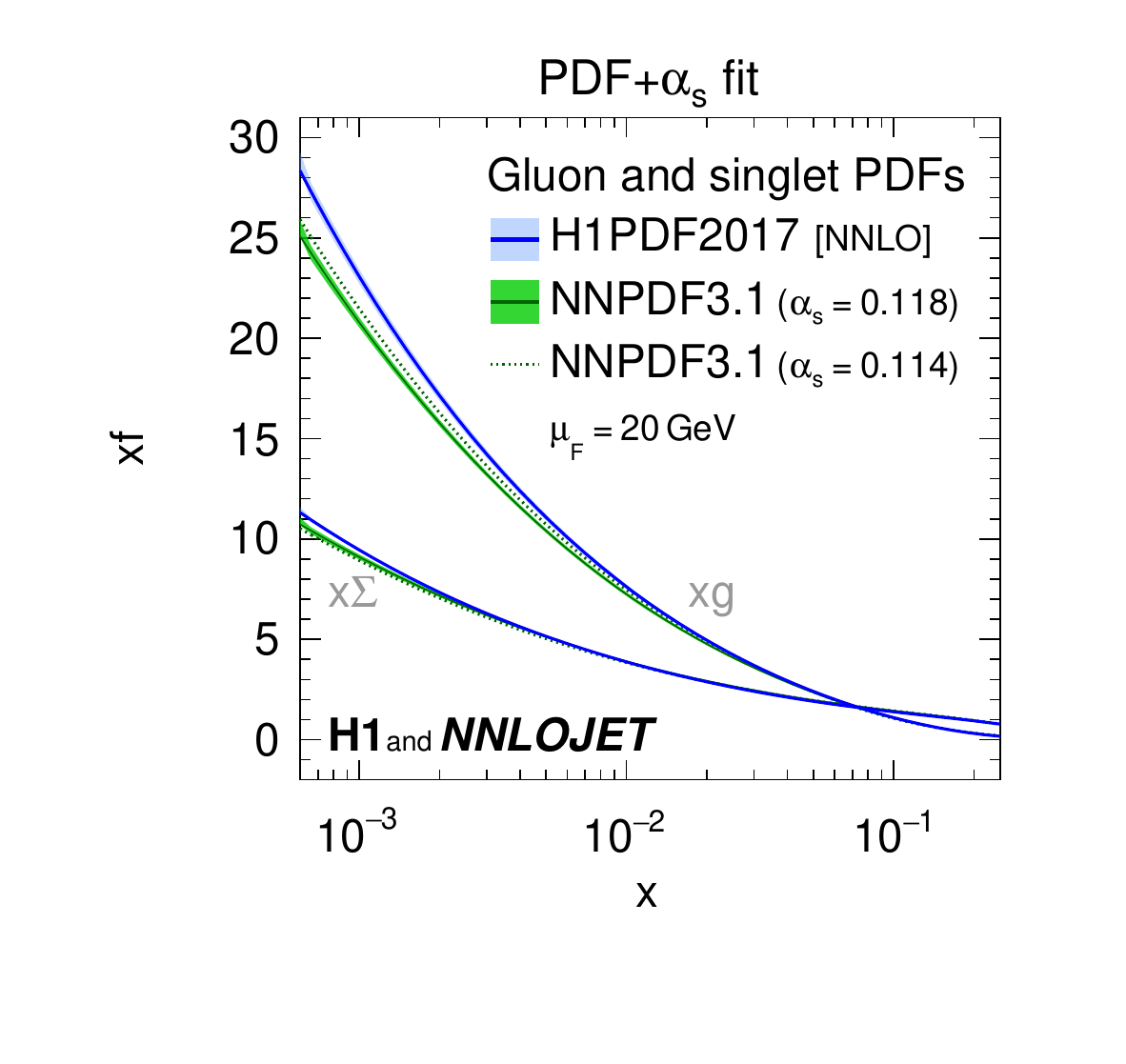}\hfill
  \hspace{-0.07\textwidth}
  \includegraphics[width=0.56\textwidth]{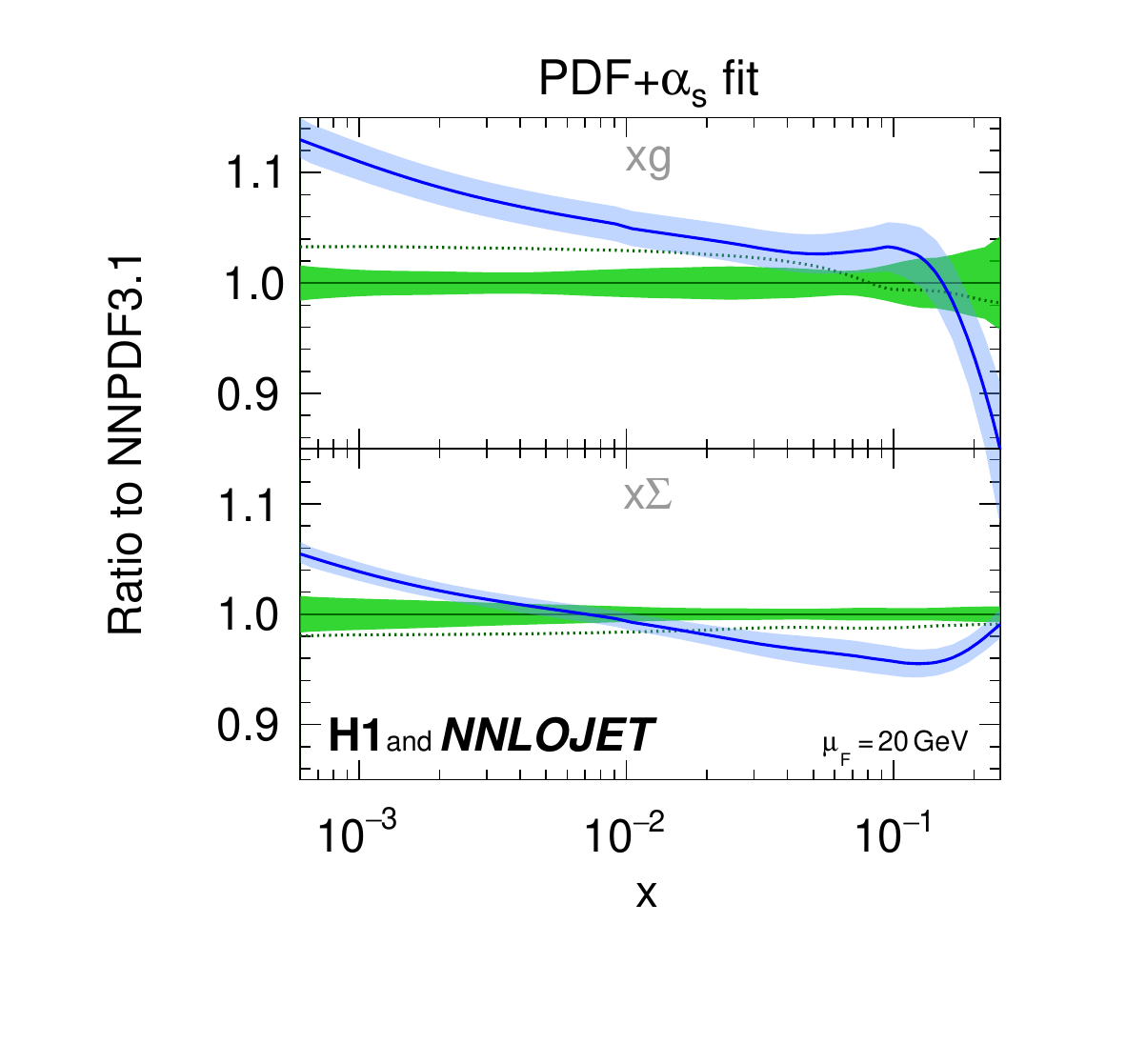}
\end{center}
\caption{
  Gluon and singlet distributions determined by the PDF+\as-fit, denoted as
  \HonePDF, as a function of
  the convolution variable $x$ (see equation~\ref{eq:sigma}).
  The distributions are displayed at $\muf=20\,\GeV$.
  The PDFs are compared to the NNPDF3.1 PDFs determined with
  values of $\asmzPDF$ of 0.114 and 0.118.
  Ratios to NNPDF3.1 are shown in the right panels.
}
\label{fig:pdfs}
\end{figure}

% --- discuss PDFs
\paragraph{PDF parametrisation results}
The PDF and \asmz\ parameters determined together in this fit (table~\ref{tab:PDFfit}) are
denoted as \HonePDF.
% LHAPDF release
It is released \cite{h1:www} in the LHAPDF \cite{Buckley:2014ana}
format with experimental, hadronisation and \asmz\ uncertainties included.
The gluon and singlet momentum distributions, $xg$ and $x\Sigma$, the latter
defined as the sum of all quark and anti-quark densities, are
compared to NNPDF3.1 at a scale 
$\muf=20\,\GeV$ in figure~\ref{fig:pdfs}.
The uncertainties of the fitted PDFs are somewhat larger
than the uncertainties of NNPDF3.1. For NNPDF3.1, \asmzPDF\ is fixed while it
is a free parameter in the \HonePDF\ fit. 
Within uncertainties, the singlet distribution obtained for \HonePDF\
is in fair agreement with NNPDF3.1 over a large range in $x$, whereas  
the gluon density is consistent with NNPDF3.1 only for $x>0.01$ and
is significantly higher than NNPDF3.1 at lower $x$.
This difference can not be explained by the 
assumptions made on the strong coupling in NNPDF3.1, as can be seen
from the NNPDF3.1 distributions obtained for
$\asmzPDF=0.114$.
However, there are differences in the datasets used for the fits.
For \HonePDF\ only H1 data are considered, restricted to the range $Q^2>10\,\text{GeV}^2$.
For NNPDF3.1 the combined HERA DIS data \cite{Abramowicz:2015mha} are
used, starting from $Q^2>3.5\,\text{GeV}^2$. Data from
other processes and experiments are also included, but no DIS jet data.

The PDFs obtained for each of the model and
parametrisation variations (not shown in figure~\ref{fig:pdfs}) 
are contained in the exp,had,PDF uncertainty band for $x>0.0004$
and thus do not
explain  the differences to NNDPF3.1.

\begin{figure}[ht]
\begin{center}
  \includegraphics[width=0.8\textwidth]{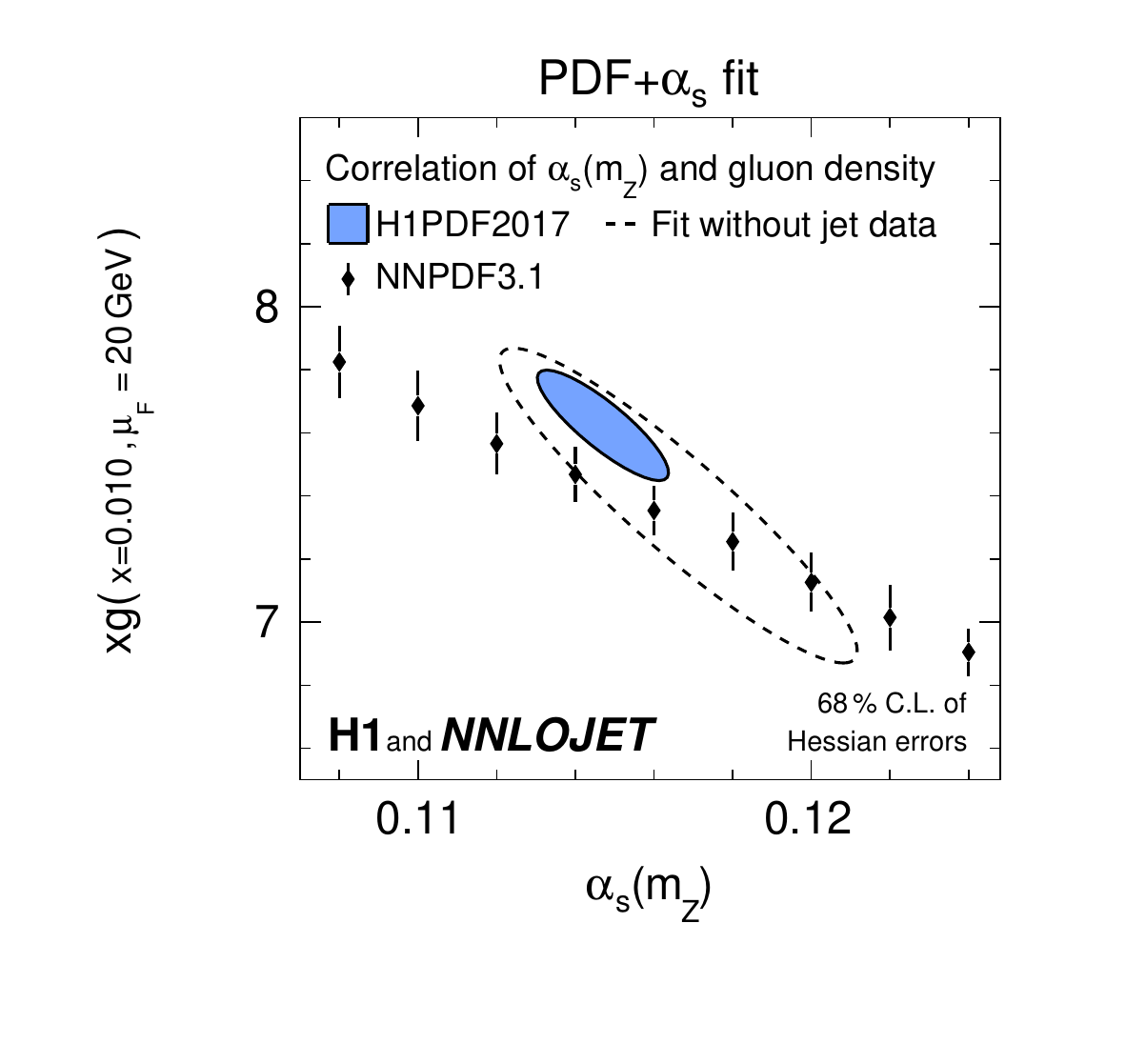}
\end{center}
\caption{
  Error ellipses of Hessian uncertainties at $68\,\%$ confidence level of \asmz\ and
  the gluon density $xg$ at $\muf=20\,\GeV$ and $x=0.01$ as a result of
  two different PDF+\as-fits.
  The filled ellipse indicates the result of the \HonePDF\ fit and the dashed
  line of a PDF+\as-fit with jet data excluded.
  The error ellipses represent the combined effect of experimental
  and hadronisation uncertainties as described in the text.
  The diamonds indicate the gluon density of the NNPDF3.1 PDF set for
  fixed values \asmzPDF.
}
\label{fig:rhoAsGluon}
\end{figure}

% --- discuss correlation gluon-alpha_s
\paragraph{The impact of H1 jet data on PDF fits}
The PDF+\as-fit is repeated with the normalised jet data excluded, i.e.\ only
inclusive DIS data are considered.
For this fit and the \HonePDF\ fit the gluon distribution $xg(x,\muf)$ is
evaluated at $\muf=20\,\GeV$ and $x=0.01$ and its Hessian uncertainty together
with its correlation coefficient with \asmz\ are calculated.
The resulting Hessian error ellipses are
displayed in figure~\ref{fig:rhoAsGluon} at a confidence level of
$68\,\%$.
Compared to the fit without jet data, the inclusion of jet data
significantly reduces the uncertainties of \asmz\ and $xg$, as well as
their correlation.
The correlation coefficient is -0.92 and reduce to
-0.85 if jet data is included.
Also shown is the gluon distribution of NNPDF3.1 determined for different values of \asmzPDF.
At this particular choice of $x$ and $\muf$, the gluon density of 
\HonePDF\ is found to be consistent with NNPDF3.1 in the range where \asmzPDF\ is close to the result of the \HonePDF\ fit.

The two fits are repeated for each of the
model and parametrisation variations (not shown in
figure~\ref{fig:rhoAsGluon}). 
For the \HonePDF\ fit, only small variations of the results are
observed, in accord with the small model and parametrisation
uncertainties assigned to \asmz{}. However, if the jet data are not
included in the fit, the resulting  \asmz\ and $xg$ are found to be
strongly dependent on the assumptions made for the PDF parametrisation.
This confirms previous observations~\cite{Adloff:2000qk}, namely that $xg$
and \asmz\ together cannot be determined reliably 
from H1 inclusive DIS data alone.

In summary, the inclusion of jet data allows for a reliable
determination of \asmz\ and its uncertainty. It also stabilises the gluon density determination.
In contrast to a previous study using only a fraction of the H1
data~\cite{Adloff:2000tq},  it can now be
stated that all H1 jet data taken together with all H1 inclusive DIS
data do allow for a simultaneous determination of $xg$ and \asmz, with
a precision on $xg$ competitive to global PDF fits obtained using fixed value
of \asmzPDF.

%%%%%%%%%%%%%%%%%%%%%%%%%%%%%%%%%%%%%%%%%%%%%%%%%%%%%%%%%%%%
%                    Summary
%%%%%%%%%%%%%%%%%%%%%%%%%%%%%%%%%%%%%%%%%%%%%%%%%%%%%%%%%%%%
%\clearpage
\section{Summary}
The new next-to-next-to-leading order pQCD calculations (NNLO) for jet
production cross sections in neutral-current DIS are exploited for a
determination of the strong coupling constant \asmz\ using inclusive
jet and dijet cross section measurements published by the H1 collaboration. 
Two methods are explored to determine the value of \asmz.

In the first approach H1 inclusive jet and dijet data are analysed.
The cross section predictions account for the \as\ dependence in the two
components of the calculations, the partonic cross sections and the parton
distribution functions (PDFs).
The strong coupling constant is determined to be
$\asmz=0.1166\,(19)_{\rm exp}\,(24)_{\rm th}$, where the
jet data are restricted to high scales $\tilmu > 28\,\GeV$.
Uncertainties due to the input PDFs or the
hadronisation corrections are found to be small, and the largest
source of uncertainty is from scale variations of the NNLO calculations.
The experimental uncertainty may be reduced to 0.8\,\%, if all
inclusive jet and dijet data with $\tilmu>2m_b$ are considered, 
but the scale uncertainties are increased significantly.
The smallest total uncertainty on \asmz\ of 2.5\,\% is obtained when
restricting the data to $\tilmu>42\,\GeV$.
Values of \asmz\ determined from inclusive jet data or dijet
data alone are found to be consistent with the main result.
All these results are found to be consistent with each other and with
the world average value of \asmz.

The running of the strong coupling constant is tested in the range of
approximately 7 to $90\,\GeV$ by dividing the jet data into ten subsets of
approximately constant scale. 
The scale dependence of the coupling is found to be consistent with
the expectation. 

In a second approach a combined determination of PDF parameters and
\asmz\ in NNLO accuracy is performed.
In this fit all normalised inclusive jet and dijet
cross sections published by H1 are analysed together with all
inclusive neutral-current and charged-current DIS cross sections
determined by H1.
Using the data with $\Qsq>10\,\GeVsq$, the value of \asmz\ is
determined to be $\asmz=\PDFasResult\,(25)_{\rm tot}$.
Consistency with the other results and the world average is found.
The resulting PDF set \HonePDF\ is found to be consistent with the
NNPDF3.1 PDF set at sufficiently large $x>0.01$, albeit there are
differences at lower $x$.
It is demonstrated that the inclusion of H1 jet data into such a simultaneous
PDF and \asmz\ determination provides stringent constraints on
\asmz\ and the gluon density.
The results and their uncertainties
are found to be largely insensitive to the assumptions made for the PDF parametrisation.

Relevant phenomenological aspects of the NNLO calculations are studied
for the first time.
The NNLO calculations are repeated for a number of different
scale choices and scale factors, as well as for a large variety of recent PDF
sets. The level of agreement with H1 jet data is judged quantitatively.
The NNLO calculations improve significantly the description of the
data and reduce the dominating theoretical uncertainty on \asmz\ in
comparison to previously employed NLO calculations.
All jet cross section measurements are found to be well described by the NNLO predictions.
These NNLO calculations are employed for a PDF determination for the first time.

This is the first precision extraction of \asmz\ from jet data
at NNLO involving a hadron in the initial state. 
It opens a new chapter of precision QCD measurements at hadron colliders.

%%%%%%%%%%%%%%%%%%%%%%%%%%%%%%%%%%%%%%%%%%%%%%%%%%%%%%%%%%%%%%%%%%%%%
%         Acknowledgements
%%%%%%%%%%%%%%%%%%%%%%%%%%%%%%%%%%%%%%%%%%%%%%%%%%%%%%%%%%%%%%%%%%%%%
%\clearpage
\section*{Acknowledgements}
We are grateful to the HERA machine group whose outstanding
efforts have made this experiment possible.
We thank the engineers and technicians for their work in constructing
and maintaining the H1 detector, our funding agencies for
financial support, the DESY technical staff for continual assistance
and the DESY directorate for support and for the
hospitality which they extend to the non--DESY
members of the collaboration.

We would like to give credit to all partners contributing to the EGI
computing infrastructure for their support for the H1 collaboration. 

We express our thanks to all those involved in securing not only the
H1 data but also the software and working environment for long term
use allowing the unique H1 data set to continue to be explored in the
coming years. The transfer from experiment specific to central
resources with long term support, including both storage and batch
systems has also been crucial to this enterprise. We therefore also
acknowledge the role played by DESY-IT and all people involved during
this transition and their future role in the years to come.

This research was supported by the Swiss National Science
Foundation (SNF) under contracts 200020-162487 and CRSII2-160814, in part by
the UK Science and Technology Facilities Council as well as by the
Research Executive Agency (REA) of the European Union under the Grant
Agreement PITN-GA-2012-316704  (``HiggsTools'') and  the ERC Advanced
Grant MC@NNLO (340983).

We gratefully express our thanks for support from the
Institute for Particle Physics Phenomenology Durham (IPPP), in the
form of an IPPP Associateship.

%%%%%%%%%%%%%%%%%%%%%%%%%%%%%%%%%%%%%%%%%%%%%%%%%%%%%%%%%%%%
%                    bib
%%%%%%%%%%%%%%%%%%%%%%%%%%%%%%%%%%%%%%%%%%%%%%%%%%%%%%%%%%%%
\clearpage
\begin{flushleft}
\bibliography{desy17-137}

\end{flushleft}

%%%%%%%%%%%%%%%%%%%%%%%%%%%%%%%%%%%%%%%%%%%%%%%%%%%%%%%%%%%%
%                    tables
%%%%%%%%%%%%%%%%%%%%%%%%%%%%%%%%%%%%%%%%%%%%%%%%%%%%%%%%%%%%
\clearpage
%\section*{Table of results}

\begin{table}[tbhp]
  \scriptsize
  \begin{center}
    \begin{tabular}{lllccc}
      \multicolumn{6}{c}{\bf\boldmath \asmz\ values from H1 jet cross sections} \\
      \hline
            {\bf Data}         %
            & {\boldmath ~~$\tilmu_{\rm cut}$}
            & \multicolumn{1}{c}{\bf\boldmath \asmz with uncertainties}
            & \multicolumn{1}{c}{\bf th}
            & \multicolumn{1}{c}{\bf tot}
            & {\boldmath $\chisq/\ndf$}
            \\     %
            \hline
            {\bf Inclusive jets}  & & & &       \\
%      {\color{red}UPDATED RESULTS from 19.Apr21}
            $300\,\GeV$ high-\Qsq & $2m_b$
            & $ 0.1253\,(33)_{\rm exp}\,(23)_{\rm had}\,(5)_{\rm PDF}\,(3)_{\rm PDF\as}\,(5)_{\rm PDFset}\,(28)_{\rm scale}$  & $(37)_{\rm th}$  & $(49)_{\rm tot}$ & $3.7/15$  %   820-HQ-IJ
            \\
            HERA-I      low-\Qsq  & $2m_b$
            & $ 0.1113\,(18)_{\rm exp}\,~\,(8)_{\rm had}\,(5)_{\rm PDF}\,(5)_{\rm PDF\as}\,(7)_{\rm PDFset}\,(33)_{\rm scale}$  & $(36)_{\rm th}$  & $(40)_{\rm tot}$ & $14.6/22$  %   H-I-LQ-IJ
            \\
            HERA-I      high-\Qsq   & $2m_b$
            & $ 0.1163\,(26)_{\rm exp}\,~\,(9)_{\rm had}\,(6)_{\rm PDF}\,(4)_{\rm PDF\as}\,(3)_{\rm PDFset}\,(22)_{\rm scale}$  & $(25)_{\rm th}$  & $(36)_{\rm tot}$ & $13.2/23$  %   H-I-HQ-IJ
            \\
            HERA-II     low-\Qsq    & $2m_b$
            & $ 0.1212\,(16)_{\rm exp}\,(12)_{\rm had}\,(4)_{\rm PDF}\,(4)_{\rm PDF\as}\,(3)_{\rm PDFset}\,(38)_{\rm scale}$  & $(40)_{\rm th}$  & $(43)_{\rm tot}$ & $28.2/40$  %   HII-LQ-IJ
            \\
            HERA-II     high-\Qsq   & $2m_b$
            & $ 0.1156\,(20)_{\rm exp}\,(10)_{\rm had}\,(5)_{\rm PDF}\,(4)_{\rm PDF\as}\,(2)_{\rm PDFset}\,(24)_{\rm scale}$  & $(27)_{\rm th}$  & $(34)_{\rm tot}$ & $33.7/29$  %   HII-HQ-IJ
            \\
            \hline
            {\bf Dijets}    & & & &    \\
            $300\,\GeV$ high-\Qsq  & $2m_b$
            & $ 0.1246\,(41)_{\rm exp}\,(18)_{\rm had}\,(5)_{\rm PDF}\,(2)_{\rm PDF\as}\,(3)_{\rm PDFset}\,(34)_{\rm scale}$  & $(39)_{\rm th}$  & $(57)_{\rm tot}$ & $8.5/15$  %   820-HQ-2J
            \\
            HERA-I      low-\Qsq   & $2m_b$
            & $ 0.1121\,(24)_{\rm exp}\,~\,(8)_{\rm had}\,(5)_{\rm PDF}\,(4)_{\rm PDF\as}\,(5)_{\rm PDFset}\,(34)_{\rm scale}$  & $(36)_{\rm th}$  & $(44)_{\rm tot}$ & $10.2/20$  %   H-I-LQ-2J
            \\
            HERA-II     low-\Qsq    & $2m_b$
            & $ 0.1198\,(12)_{\rm exp}\,(12)_{\rm had}\,(5)_{\rm PDF}\,(5)_{\rm PDF\as}\,(3)_{\rm PDFset}\,(42)_{\rm scale}$  & $(44)_{\rm th}$  & $(45)_{\rm tot}$ & $17.0/41$  %   HII-LQ-2J
            \\
            HERA-II     high-\Qsq   & $2m_b$
            & $ 0.1116\,(22)_{\rm exp}\,~\,(7)_{\rm had}\,(5)_{\rm PDF}\,(3)_{\rm PDF\as}\,(3)_{\rm PDFset}\,(15)_{\rm scale}$  & $(18)_{\rm th}$  & $(29)_{\rm tot}$ & $21.5/23$  %   HII-HQ-2J
            \\
            \hline
            H1 inclusive jets  & $2m_b$
  & $ 0.1157\,(10)_{\rm exp}\,~\,(6)_{\rm had}\,(4)_{\rm PDF}\,(4)_{\rm PDF\as}\,(2)_{\rm PDFset}\,(34)_{\rm scale}$  & $(36)_{\rm th}$  & $(37)_{\rm tot}$ & $118.1/133$  %   IJ
            \\
            H1 inclusive jets & $28\,\GeV$
  & $ 0.1158\,(19)_{\rm exp}\,~\,(9)_{\rm had}\,(2)_{\rm PDF}\,(2)_{\rm PDF\as}\,(4)_{\rm PDFset}\,(21)_{\rm scale}$  & $(23)_{\rm th}$  & $(30)_{\rm tot}$ & $43.0/60$  %   IJ28
            \\
            H1 dijets  & $2m_b$
  & $ 0.1174\,(11)_{\rm exp}\,~\,(8)_{\rm had}\,(5)_{\rm PDF}\,(4)_{\rm PDF\as}\,(3)_{\rm PDFset}\,(33)_{\rm scale}$  & $(36)_{\rm th}$  & $(38)_{\rm tot}$ & $80.3/102$  %   2J
            \\
            H1 dijets & $28\,\GeV$
  & $ 0.1157\,(22)_{\rm exp}\,(12)_{\rm had}\,(3)_{\rm PDF}\,(2)_{\rm PDF\as}\,(3)_{\rm PDFset}\,(19)_{\rm scale}$  & $(23)_{\rm th}$  & $(32)_{\rm tot}$ & $31.6/43$  %   2J28
            \\
            \hline
            H1 jets & $2m_b$
  & $ 0.1170\,~\,(9)_{\rm exp}\,~\,(7)_{\rm had}\,(5)_{\rm PDF}\,(4)_{\rm PDF\as}\,(2)_{\rm PDFset}\,(38)_{\rm scale}$  & $(39)_{\rm th}$  & $(40)_{\rm tot}$ & $173.0/199$  %   MJ
            \\
            H1 jets & $28\,\GeV$
  & $ 0.1166\,(19)_{\rm exp}\,~\,(9)_{\rm had}\,(3)_{\rm PDF}\,(2)_{\rm PDF\as}\,(4)_{\rm PDFset}\,(21)_{\rm scale}$  & $(24)_{\rm th}$  & $(30)_{\rm tot}$ & $62.4/90$  %   MJ28
            \\
            H1 jets   & $42\,\GeV$
  & $ 0.1172\,(23)_{\rm exp}\,~\,(8)_{\rm had}\,(2)_{\rm PDF}\,(2)_{\rm PDF\as}\,(7)_{\rm PDFset}\,(14)_{\rm scale}$  & $(18)_{\rm th}$  & $(29)_{\rm tot}$ & $37.0/40$  %   MJ42
            \\
            \hline
             \HonePDF & $2m_b$
            % /nfs/dust/h1/group/britzger/alpos/Alpos/nnlo_alphas/fix2020_pdffit_nnlojetgrids/log.12.4PDF.ErrorDef0p01.txt -> 4th fit!
            &  $\PDFasResult\,(11)_{\rm exp,NP,PDF}\,(2)_{\rm mod}\,(3)_{\rm par}$\hfill$\,(23)_{\rm scale}$
            &
           & $(25)_{\rm tot}$
           & $1518.6/1516$
            %           & $\PDFasResult\,(11)_{\rm exp,NP,PDF}\,(2)_{\rm mod}\,(2)_{\rm par}\,(26)_{\rm scale}$
%           &
%           & $(28)_{\rm tot}$
%           & $1539.7/1516$
            \\
            \hline
    \end{tabular}
    \caption{Summary of values of \asmz\ from fits to H1 jet cross
      section measurements using NNLO predictions.
      The uncertainties denote the experimental (exp), hadronisation
      (had), PDF, PDF\as, PDFset and scale uncertainties as described
      in the text.
      The rightmost three columns denote the quadratic sum of the
      theoretical uncertainties (th), the total (tot) uncertainties
      and the value of $\chisq/\ndf$ of the corresponding fit.
      Along the vertical direction, the table data are segmented into five
      parts. The uppermost part summarises fits to individual
      inclusive jet datasets. The second part corresponds to fits of the
      individual dijet datasets. The third part summarises fits to all
      inclusive jets or all dijets together, with different choices of
      the lower cut on the scale $\tilmu_{\rm cut}$. The fourth group
      of fits, labelled H1 jets, is made using all available dijet and
      inclusive jet data together, for three different choices of
      $\tilmu_{\rm cut}$. The bottom row corresponds to a combined fit
      of inclusive data and normalised jet data. For that fit,
      theoretical uncertainties related to the PDF determination
      interfere with the experimental uncertainties and thus no overall
      theoretical uncertainty is quoted.
    }
    \label{tab:asresults}
    \end{center}
\end{table}

% ----- table with all alpha_s values
\begin{table}[tbhp]
  %\footnotesize
  \scriptsize
  \begin{center}

    \begin{tabular}{cc@{\hskip5pt}cc@{\hskip5pt}cc@{\hskip5pt}c}
      \multicolumn{7}{c}{\bf\boldmath Running of the strong coupling} \\
      \hline
            {\bf \mur}         %
            & \multicolumn{2}{c}{Inclusive jets}
            & \multicolumn{2}{c}{Dijets}
            & \multicolumn{2}{c}{H1 jets}
            \\
            $[$GeV$]$
            & \multicolumn{1}{c}{\asmz}
            &  \multicolumn{1}{c}{\asmur}
            & \multicolumn{1}{c}{\asmz}
            &  \multicolumn{1}{c}{\asmur}
            & \multicolumn{1}{c}{\asmz}
            &  \multicolumn{1}{c}{\asmur}
            \\     %
            \hline
 & $0.1170\,(12)\,(41)$   & $ 0.1909\,(33)\,(119)$ % $0.18--\,(34)\,(114)  $  %   IJ  mu1   
 & $0.1207\,(25)\,(40)$   & $ 0.1969\,(70)\,(113)$ % $0.19--\,(77)\,(116)  $  %   2J  mu1   
 & $0.1170\,(12)\,(42)$   & $ 0.1890\,(33)\,(116)$ % $0.18--\,(34)\,(114)  $  %   MJ  mu1   
            \\
            10.1 % 10.198. Now: 10.1193                                                     
 & $ 0.1161\,(17)\,(35)$   & $ 0.1750\,(40)\,(90) $ %   $0.16--\,(39)\,(81)  $  %   IJ  mu2 
 & $ 0.1192\,(14)\,(42)$   & $ 0.1785\,(32)\,(99) $ %   $0.17--\,(34)\,(99)  $  %   2J  mu2 
 & $ 0.1173\,(13)\,(38)$   & $ 0.1761\,(30)\,(90) $ %   $0.17--\,(31)\,(91)  $  %   MJ  mu2 
            \\
            13.3 % 13.2665                                                                  
 & $ 0.1167\,(15)\,(41)$   & $ 0.1663\,(31)\,(84) $ %  $0.16--\,(30)\,(88)  $  %   IJ  mu3  
 & $ 0.1152\,(18)\,(35)$   & $ 0.1598\,(36)\,(70) $ %  $0.15--\,(36)\,(76)  $  %   2J  mu3  
 & $ 0.1165\,(15)\,(40)$   & $ 0.1641\,(31)\,(82) $ %  $0.16--\,(30)\,(86)  $  %   MJ  mu3  
            \\
            17.2 % 17.2337                                                                  
 & $ 0.1160\,(15)\,(29)$   & $ 0.1560\,(28)\,(52) $ %  $0.14--\,(26)\,(59)  $  %   IJ  mu4  
 & $ 0.1136\,(19)\,(24)$   & $ 0.1488\,(33)\,(42) $ %  $0.14--\,(33)\,(53)  $  %   2J  mu4  
 & $ 0.1158\,(16)\,(28)$   & $ 0.1541\,(29)\,(51) $ %  $0.14--\,(27)\,(59)  $  %   MJ  mu4  
            \\
            20.1 % 20.1246                                                                  
 & $ 0.1158\,(18)\,(28)$   & $ 0.1509\,(31)\,(49) $ %  $0.14--\,(29)\,(56)  $  %   IJ  mu5  
 & $ 0.1139\,(21)\,(26)$   & $ 0.1450\,(35)\,(43) $ %  $0.14--\,(36)\,(52)  $  %   2J  mu5  
 & $ 0.1156\,(17)\,(28)$   & $ 0.1492\,(29)\,(48) $ %  $0.14--\,(29)\,(55)  $  %   MJ  mu5  
            \\
            24.5 % 24.4949                                                                  
 & $ 0.1184\,(16)\,(28)$   & $ 0.1495\,(26)\,(44) $ %  $0.14--\,(26)\,(48)  $  %   IJ  mu6  
 & $ 0.1153\,(22)\,(22)$   & $ 0.1419\,(34)\,(34) $ %  $0.14--\,(36)\,(38)  $  %   2J  mu6  
 & $ 0.1182\,(17)\,(27)$   & $ 0.1478\,(27)\,(43) $ %  $0.14--\,(27)\,(46)  $  %   MJ  mu6  
            \\
            29.3 % 29.3258                                                                  
 & $ 0.1091\,(32)\,(31)$   & $ 0.1307\,(47)\,(48) $ %  $0.12--\,(51)\,(41)  $  %   IJ  mu7  
 & $ 0.1174\,(26)\,(36)$   & $ 0.1404\,(38)\,(52) $ %  $0.14--\,(50)\,(50)  $  %   2J  mu7  
 & $ 0.1142\,(25)\,(33)$   & $ 0.1370\,(37)\,(48) $ %  $0.13--\,(44)\,(46)  $  %   MJ  mu7  
            \\
            36.0 % 35.9166                                                                  
 & $ 0.1164\,(27)\,(38)$   & $ 0.1364\,(38)\,(49) $ %  $0.13--\,(43)\,(50)  $  %   IJ  mu8  
 & $ 0.1146\,(31)\,(32)$   & $ 0.1317\,(41)\,(43) $ %  $0.13--\,(50)\,(39)  $  %   2J  mu8  
 & $ 0.1156\,(26)\,(35)$   & $ 0.1342\,(35)\,(48) $ %  $0.13--\,(41)\,(44)  $  %   MJ  mu8  
            \\
            49.0 % 48.9898                                                                  
 & $ 0.1174\,(22)\,(19)$   & $ 0.1306\,(27)\,(22) $ %  $ 0.12--\,(27)\,(25) $  %   IJ  mu9  
 & $ 0.1134\,(31)\,(14)$   & $ 0.1237\,(37)\,(17) $ %  $ 0.12--\,(37)\,(18) $  %   2J  mu9  
 & $ 0.1174\,(23)\,(18)$   & $ 0.1295\,(28)\,(22) $ %  $ 0.12--\,(28)\,(24) $  %   MJ  mu9  
            \\
            77.5 % 77.4597                                                                  
 & $ 0.1082\,(51)\,(22)$   & $ 0.1115\,(54)\,(22) $ %  $0.11--\,(58)\,(20)  $  %   IJ  mu10 
 & $ 0.1045\,(77)\,(19)$   & $ 0.1061\,(80)\,(20) $ %  $0.10--\,(88)\,(20)  $  %   2J  mu10 
 & $ 0.1083\,(51)\,(21)$   & $ 0.1107\,(53)\,(22) $ %  $0.11--\,(58)\,(20)  $  %   MJ  mu10
            \\
            \hline            
    \end{tabular}
    \caption{
%      {\color{red}Left values(are all updated 19.Apr21. Right values also updated}
      Values of the strong coupling constant \asmur\ and at the $Z$-boson
      mass, \asmz, obtained from fits to groups of data points with
      comparable values of $\mur$.
      The first (second) uncertainty of each point corresponds to the
      experimental (theory) uncertainty. The theory uncertainties include
      PDF related uncertainties and the dominating scale uncertainty.
    }
    \label{tab:running}
    \end{center}
\end{table}

% ----- table with result of PDF fit
\begin{table}[tbhp]
  \scriptsize
  \begin{center}
    \begin{tabular}{c@{\hskip4pt}r@{$\,\pm\,$}l@{\hskip8pt}c@{\hskip4pt}c@{\hskip4pt}c@{\hskip4pt}c@{\hskip4pt}c@{\hskip4pt}c@{\hskip4pt}c@{\hskip4pt}c@{\hskip4pt}c@{\hskip4pt}c@{\hskip4pt}c@{\hskip4pt}c@{\hskip4pt}c@{\hskip2pt}}
      \multicolumn{16}{c}{\boldmath\bf Results for the PDF+{\as}-fit} \\
      \hline
      Parameter  & \multicolumn{2}{c}{Fit result}          &
      \multicolumn{13}{c}{Correlation coefficients} \\
                 & \multicolumn{2}{c}{}                    & \asmz & $g_B$ & $g_C$ & $g_D$ & $\tilde{u}_B$ & $\tilde{u}_C$ & $\tilde{u}_E$ & $\tilde{d}_B$ & $\tilde{d}_C$ & $\bar{U}_C$ & $\bar{D}_A$ & $\bar{D}_B$ & $\bar{D}_C$ \\
      \hline
      \asmz          & $\PDFasResult$& $0.0011 $  &   1 \\
      $g_B$          & $ -0.030    $ & $0.029  $  & $\hphantom{-}0.54$ &  1  \\
      $g_C$          & $    4.34   $ & $0.82   $  & $\hphantom{-}0.08$ & $\hphantom{-}0.49$ &  1 \\
      $g_D$          & $  -1.91    $ & $0.36   $  & $\hphantom{-}0.30$ & $\hphantom{-}0.25$ & $\hphantom{-}0.81$ &  1 \\
      $\tilde{u}_B$  & $    0.721  $ & $0.025  $  & $\hphantom{-}0.34$ & $\hphantom{-}0.50$ & $\hphantom{-}0.10$ & $\hphantom{-}0.05$ &  1 \\
      $\tilde{u}_C$  & $    4.904  $ & $0.081  $  & $           -0.19$ & $         {-}0.20$ & $         {-}0.17$ & $         {-}0.16$ & $         {-}0.67$&  1 \\
      $\tilde{u}_E$  & $    12.3   $ & $1.4    $  & $\hphantom{-}0.04$ & $         {-}0.31$ & $         {-}0.23$ & $         {-}0.01$ & $         {-}0.10$ & $\hphantom{-}0.62$ &  1 \\
      $\tilde{d}_B$  & $    1.032  $ & $0.080  $  & $\hphantom{-}0.16$ & $\hphantom{-}0.18$ & $\hphantom{-}0.20$ & $         {-}0.11$ & $\hphantom{-}0.22$ & $         {-}0.27$ & $         {-}0.11$ &  1 \\
      $\tilde{d}_C$  & $    5.27   $ & $0.44   $  & $           -0.13$ & $\hphantom{-}0.00$ & $         {-}0.05$ & $         {-}0.04$ & $         {-}0.18$ & $         {-}0.22$ & $\hphantom{-}0.03$ &  $\hphantom{-}0.80$ &  1 \\
      $\bar{U}_C$    & $    4.51   $ & $0.50   $  & $\hphantom{-}0.21$ & $\hphantom{-}0.03$ & $         {-}0.04$ & $\hphantom{-}0.09$ & $\hphantom{-}0.61$ & $\hphantom{-}0.07$ & $         {-}0.22$ &  $\hphantom{-}0.28$ &  $         {-}0.04$ &  1 \\
      $\bar{D}_A$    & $   0.268   $ & $0.016  $  & $\hphantom{-}0.27$ & $         {-}0.41$ & $         {-}0.16$ & $\hphantom{-}0.13$ & $\hphantom{-}0.11$ & $         {-}0.01$ & $\hphantom{-}0.10$ &  $\hphantom{-}0.11$ &  $         {-}0.04$ &  $\hphantom{-}0.44$ &  1 \\
      $\bar{D}_B$    & $ -0.107$     & $0.010  $  & $\hphantom{-}0.14$ & $         {-}0.59$ & $         {-}0.21$ & $\hphantom{-}0.08$ & $         {-}0.08$ & $\hphantom{-}0.02$ & $\hphantom{-}0.18$ &  $         {-}0.00$ &  $         {-}0.07$ &  $\hphantom{-}0.29$ &  $\hphantom{-}0.91$ &  1 \\
      $\bar{D}_C$    & $    13.09$   & $0.22    $ & $           -0.18$ & $         {-}0.31$ & $         {-}0.41$ & $         {-}0.17$ & $         {-}0.12$ & $         {-}0.10$ & $         {-}0.07$ &  $\hphantom{-}0.20$ &  $\hphantom{-}0.26$ &  $         {-}0.15$ &  $\hphantom{-}0.20$ &  $\hphantom{-}0.18$ &  1 \\
      \hline
      $g_A$          & \multicolumn{1}{r@{$\,~~\,\,$}}{3.35} &  & \multicolumn{13}{c}{constrained by sum-rules} \\
      $\tilde{u}_A$  & \multicolumn{1}{r@{$\,~~\,\,$}}{4.29} &  & \multicolumn{13}{c}{constrained by sum-rules} \\
      $\tilde{d}_A$  & \multicolumn{1}{r@{$\,~~\,\,$}}{6.78} &  & \multicolumn{13}{c}{constrained by sum-rules} \\
      $\bar{U}_A$    & \multicolumn{1}{r@{$\,~~\,\,$}}{0.161} &  & \multicolumn{13}{c}{set equal to $\bar{D}_A(1-f_s)$} \\
      $\bar{U}_B$    & \multicolumn{1}{r@{$\,~~\,\,$}}{$-0.107$} &  & \multicolumn{13}{c}{set equal to $\bar{D}_B$} \\
      \hline
    \end{tabular}
    \caption{
      Results of the PDF+{\as} fit.
      The columns denote the resulting fit value, its uncertainty and the correlations to the other parameters.
    }    
    \label{tab:PDFfit}
    \end{center}
\end{table}

%%%%%%%%%%%%%%%%%%%%%%%%%%%%%%%%%%%%%%%%%%%%%%%%%%%%%%%%%%%%
%                    figures
%%%%%%%%%%%%%%%%%%%%%%%%%%%%%%%%%%%%%%%%%%%%%%%%%%%%%%%%%%%%
\clearpage

%%%%%%%% Fig.:  %%%%%%%%%%%%%%%%%%%%

%%%%%%%% Fig.:  %%%%%%%%%%%%%%%%%%%%

%%%%%%%% Fig.:  %%%%%%%%%%%%%%%%%%%%

%%%%%%%% Fig.:  %%%%%%%%%%%%%%%%%%%%

%%%%%%%% Fig.:  %%%%%%%%%%%%%%%%%%%%

%%%%%%%% Fig.:  %%%%%%%%%%%%%%%%%%%%

%%%%%%%% Fig.:  %%%%%%%%%%%%%%%%%%%%

%%%%%%%% Fig.:  %%%%%%%%%%%%%%%%%%%%

%%%%%%%% Fig.:  %%%%%%%%%%%%%%%%%%%%

%%%%%%%% Fig.:  %%%%%%%%%%%%%%%%%%%%

%%%%%%%% Fig.: running %%%%%%%%%%%%%%%%%%%%

%%%%%%%% Fig.: cross sections  %%%%%%%%%%%%%%%%%%%%

%%%%%%%% Fig.: running %%%%%%%%%%%%%%%%%%%%

%%%%%%%% Fig.: running %%%%%%%%%%%%%%%%%%%%

%%%%%%%% Fig.:  PDFfits %%%%%%%%%%%%%%%%%%%%

%%%%%%%% Fig.:  PDFs: gluon-singlet %%%%%%%%%%%%%%%%%%%%

%%%%%%%% Fig.:  PDFs: correlation %%%%%%%%%%%%%%%%%%%%

\end{document}